\documentclass{article}

\usepackage{amsmath,amssymb,amsfonts,dcolumn,color,graphicx,graphics,latexsym,placeins,epsfig}
\usepackage{graphicx}  
\usepackage{amsmath}   
\usepackage{amssymb}   
\usepackage{bm} 
\usepackage{dcolumn}
\usepackage{color}
\usepackage{footmisc}
\usepackage{mathrsfs}
\usepackage{amsfonts}
\usepackage{varioref}
\usepackage{slashed}
\usepackage{footnote}
\usepackage[belowskip=-17pt,aboveskip=15pt]{caption}
\usepackage{amsmath}
\usepackage{makeidx}
\usepackage{amssymb}
\usepackage{graphicx}
\usepackage{bmpsize}
\usepackage{multicol}
\usepackage{url}
\usepackage{ragged2e}

\usepackage[margin=1.5in]{geometry}
\usepackage{graphicx}
\usepackage{dcolumn}
\usepackage{amsmath,amssymb,mathtools}
\usepackage{epstopdf}
\usepackage{verbatim}
\usepackage[colorlinks=true, citecolor=blue, urlcolor = blue, linkcolor= blue, 
bookmarks=true]{hyperref}
\usepackage[utf8]{inputenc}
\usepackage{amsmath}
\def\e{\epsilon}
\def\p{\partial}

\def\be{\begin{eqnarray}}
\def\ee{\end{eqnarray}}

\def\e{\epsilon}
\def\p{\partial}
\labelformat{section}{Section #1} 
\labelformat{subsection}{Section #1} 
\labelformat{subsubsection}{Section #1}
\labelformat{subsubsubsection}{Section #1}
\labelformat{eqnarray}{Eq.~(#1)} 
\labelformat{figure}{Fig.~#1} 
\labelformat{subfigure}{Fig.~\thefigure#1} 
\labelformat{table}{Tab.~#1} 
\labelformat{appendix}{Appendix #1}
\vspace{25pt}
\title{\bf Scattering of massive spin-2 field via graviton exchanges with different spin fields and the long range gravitational potential}
\author{\bf Avijit Sen Majumder\footnote{senmajumderavijit@gmail.com}~           
          and  Sourav Bhattacharya\footnote{sbhatta.physics@jadavpuruniversity.in}\\
\small{Relativity and Cosmology Research Centre, Department of Physics, Jadavpur University, Kolkata 700 032, India}\\}

\begin{document}

\maketitle

\begin{abstract}
\noindent
In this work, we compute the graviton mediated scattering amplitude of a massive spin-2 Fierz-Pauli field with various other massive spin fields, and in the non-relativistic limit, find out the corresponding two-body gravitational potentials. The massive spin-2  field does not represent gravity here. The theory of gravity is taken to be the usual massless general relativity, and the massive spin-2 field is taken as a test quantum field coupled to gravity via the standard minimal prescription. We first compute the tree level 2-2 scattering of a massive spin-2 field with massive scalar,  spin-1, and spin-1/2 fields with one graviton exchanges. Leading Newton potential, as well as the subleading spin or polarisation dependent terms at ${\cal O}(G)$ have been computed. We also consider the next to the leading order (${\cal O}(G^2)$) scattering of the massive spin-2 field with a massive scalar,  and demonstrate the spin independent, spherically symmetric leading part of the two body gravitational potential. The present paper can be considered as an attempt to compute the gravitational potential in the context of a higher spin field theory.
\end{abstract}
\section{Introduction} \label{S1}
 
 One of the most challenging unsolved tasks of modern physics is to unify the four fundamental  forces of nature. Quantising gravity is supposed to be an  integral part of this programme. However, it is well known for long that gravity is not renormalisable perturbatively when treated as a quantum field~\cite{Utiyama, Weinberg:1965nx, Boulware:1972yco, tHooft:1974toh, deser1, deser2, deser3, Stelle, Voronov, Goroff:1985th, PRD:87, Lavrov}, suggesting there must be some alternative, or more fundamental description of it which remains hitherto unknown. Despite this drawback, there has been a lot of effort given by the community to investigate such a perturbative quantum gravity framework at the first few orders and to produce physical predictions, with the hope that they will be testable in a not too far away future. Most importantly, such predictions, if or when tested, can tell us whether  the {\it graviton}, i.e., a massless spin-2 particle   as the quantum of gravity, makes any sense at all. If it indeed does so, one may also expect to  get insight about extending this perturbative framework  into  a   bigger and complete theory of quantum gravity. We refer our reader to e.g.~\cite{Mukhanov, Woodard:2014jba, Keifer, Shapiro} and references therein for various discussions and computations on such aspects.

A very interesting and physically well motivated such perturbative computation is to investigate  the gravitational potential  between two particles, via the non-relativistic limit of the 2-2 scattering amplitude with graviton exchanges. We refer our reader to~\cite{Pagels:1966zza, Iwasaki:1971vb, Hiida:1972xs, Barker:1975ae, Donoghue:1993eb, Donoghue:1994dn, Muzinich:1995uj, Hamber:1995cq, Bjerrum-Bohr:2002fji, Bjerrum-Bohr:2002gqz, deBrito:2020wmp, Majumder:2025gou, Arbuzov:2021yai, Latosh:2023zsi, Latosh:2024lhl} and references therein for such computations for massive spin-0, spin-1/2 and spin-1 fields. For fields with spin, polarisation or spin dependence of the gravitational potential were computed. We further refer or reader to~\cite{Donoghue:1995cz, Akhundov:1996jd, Bjerrum-Bohr:2002aqa, Donoghue_1996, MODANESE1995697, Bern:2002kj, Burgess:2003jk, Goldberger:2004jt, Akhundov:2006gh, Holstein:2008sx, Bjerrum-Bohr:2014zsa, Bjerrum-Bohr:2015vda, Bai:2016ivl, Toms:2010vy, Malta2015ComparativeAO, Frob:2016xte, Foffa:2016rgu, Foffa:2016rgu, Donoghue:2017pgk, Ulhoa_2017, Levi:2018nxp, Olyaei:2018asy, Frob:2021mpb} and references therein for various computations and predictions on perturbative quantum gravity, including computation of gravitational potential and bending of light using the effective field theory framework at various post-Minkowski  orders.

In this paper, we wish to compute the scattering of a massive spin-2 Fierz-Pauli field~\cite{Fierz Pauli original, Bekenstein:1972ky, Dvali:2006az, deRham:2010gu, deRham:2010ik, Folkerts:2013mra, Koenigstein_2016, Jalali:2020dfj, Gill:2023kyz, Farolfi:2025knq} with various other spin fields, and compute the two-body gravitational potential from the non-relativistic limit of the 2-2 scattering amplitude. We will take the gravity to be Einstein's general relativity, with the usual massless spin-2 graviton as its quanta. In other words, the gravity will {\it not} be taken to be massive here. The massive spin-2 field will be taken as a test field minimally and covariantly coupled to Einstein's gravity.  The present analysis should be interpreted as probing a spin-2 field as a massive particle, analogous to the Proca (massive spin-1) field. In other words, along with the usual graviton, we are  considering another spin-2 field here as an independent quantum field. The chief physical motivation behind such a study is to understand the perturbative quantum gravity effect in the presence of a massive higher spin field theory. Indeed, such consideration of a massive spin-2 meson field is not new, and can be seen in  earlier references, e.g.~\cite{Bekenstein:1972ky} in the context of black hole no hair theorems. See also~\cite{Dvali:2006az} for a treatment of a black hole as a massive spin-2 condensate. It was argued 
in~\cite{Folkerts:2013mra} that any massive spin-2 field theory cannot be unique. In~\cite{Jalali:2020dfj}, the Hamiltonian structure of the Pauli-Fierz theory can be seen in curved spacetimes. Graviton-photon production in the presence of a massive spin-2 field can be seen in~\cite{Gill:2023kyz}. Finally, we also refer our reader to  e.g.~\cite{vanDam:1970vg, Zakharov, Nicolis, deRham:2014zqa, Gambuti:2021meo, Tachinami:2024iqg} and references therein for various aspects of massive gravity theory.

Thus, this paper can be taken as an attempt to compute the gravitational potential in the context of a higher spin field theory, which happens to be the Fierz-Pauli field presently.
The rest of the paper is organised as follows. In the next section, we briefly review the basic setup we will be working in. In~\ref{S3} we compute the tree level 2-2 scattering of a massive spin-2 field, respectively with massive spin-0, spin-1, and spin-1/2 fields via one graviton exchanges. In the non-relativistic limit of the scattering amplitude, we will compute the tree level  ${\cal O}(G)$ gravitational potential. Various spin and polarisation dependent terms will  be explicitly demonstrated. We further extend the massive spin-2-spin-0 scattering to ${\cal O}(G^2)$ in \ref{S4}. We explicitly demonstrate only the polarisation independent, leading spherically symmetric parts of the potential in this case. Finally, we conclude in \ref{S5}.  We will work with the mostly positive signature of the metric in $3+1$-dimensions and will set $c=\hbar=1$. We will also use the notation  for symmetrisation : $X_{(\mu \nu)}= X_{\mu\nu}+X_{\nu\mu}$.

\bigskip
\section{The basic ingredients}\label{S2}

We now wish to outline the basic setup we will be working in. The general, minimally coupled  action reads
\begin{equation}
S= \int d^4 x \sqrt{-g} \left[ \frac{2 R}{\kappa^2}+ {\cal L}_M\right]
\label{qg0}
\end{equation}
where $\kappa^2=32\pi G$ and ${\cal L}_M$ collectively represent  the matter fields. Perturbative expansions around the flat or Minkowski background  for our purpose read,
\begin{equation}
    \begin{split}
        & g_{\mu\nu} = \eta_{\mu\nu} +\kappa h_{\mu\nu};\, \,  g^{\mu\nu}= \eta^{\mu\nu} -\kappa h^{\mu\nu} +\kappa^2 h^{\mu}{}_{\alpha} h^{\alpha \nu}+{\cal O}(\kappa^3); \,\,\,\,\,
         \sqrt{-g} = 1+\frac{\kappa h}{2}+\frac{\kappa^2 h^2}{8} - \frac{\kappa^2}{4} h_{\mu\nu}h^{\mu\nu} + {\cal O}(\kappa^3) \\
        &   \Gamma^{\mu}_{\nu\rho} = \frac{\kappa}{2}\eta^{\mu\alpha}\left(\p_{\nu}h_{\rho\alpha}+\p_{\rho} h_{\nu\alpha}- \p_{\alpha} h_{\nu\rho} \right)-\frac{\kappa^2}{2}h^{\mu\alpha}(\p_{\nu}h_{\rho\alpha} +\p_{\rho}  h_{\nu\alpha}- \p_{\alpha} h_{\nu\rho} )+ \frac{\kappa^3}{2}h^{\mu\beta}h_{\beta}{}^{\alpha} \left(\p_{\nu}h_{\rho\alpha}+\p_{\rho} h_{\nu\alpha}- \p_{\alpha} h_{\nu\rho} \right) +{\cal O}(\kappa^4)
        \label{qg17}
    \end{split}
\end{equation}
\noindent
In the de Donder gauge, 
\begin{equation}
    \p_{\mu}\Big(h^{\mu\nu}-\frac12h \eta^{\mu\nu}  \Big)=0,
\label{qg12}
\end{equation}
the free graviton propagator, appropriate for the mostly positive metric signature reads
\begin{equation}
\Delta_{\mu\nu\alpha\beta}(k)= - \frac{i {\cal P}_{\mu\nu\alpha\beta}}{k^2} 
\label{qg4}
\end{equation}
where 
\begin{equation}
    {\cal P}_{\mu\nu\alpha\beta}:= \frac12\left(\eta_{\mu\alpha}\eta_{\nu\beta}+\eta_{\mu\beta}\eta_{\nu\alpha}-\eta_{\mu\nu}\eta_{\alpha\beta}\right).
\end{equation}
%
\subsection{The massive spin-2 setup}
Following \cite{Fierz Pauli original, Bekenstein:1972ky,  Dvali:2006az, deRham:2010ik, Folkerts:2013mra, Koenigstein_2016, Jalali:2020dfj,  Gill:2023kyz, Farolfi:2025knq,  Gambuti:2021meo, Tachinami:2024iqg}, we now wish to review very briefly the basic structure of a massive spin-2 field theory coupled to gravity. The general action  reads
\begin{equation}
\begin{split}
      S_{\text{massive spin-2}} = &\int d^4x \sqrt{-g}\Big[- \frac12 (\nabla_{\lambda}H^{\mu\nu})(\nabla^{\lambda}H_{\mu\nu})+(\nabla^{\lambda}H_{\mu\lambda})(\nabla_{\alpha}H^{\mu\alpha}) -(\nabla_{\lambda}H)(\nabla_{\mu}H^{\mu\lambda}) \\
      & + \frac12 (\nabla_{\lambda}H)(\nabla^{\lambda}H)  -\frac12 M^2 \Big(H_{\mu\nu}H^{\mu\nu} -H^2 \Big)\Big]
\label{ml1}
\end{split}
\end{equation} 
where $H_{\mu\nu}$ is the symmetric Fierz-Pauli massive spin-2 field, all the contractions are done with respect to  $g_{\mu\nu}$, with $H=g_{\mu\nu}H^{\mu\nu}$.  `$\nabla$' is the covariant derivative defined with respect to the metric $g_{\mu\nu}=\eta_{\mu\nu}+\kappa h_{\mu\nu}$. The equation of motion corresponding to  Eq.~\ref{ml1} in the flat background reads
\begin{equation}
\p^2 H_{\mu\nu} - \p_{(\nu}\p^{\alpha}H_{\mu)\alpha}+\eta_{\mu\nu} \p_{\alpha}\p_{\beta} H^{\alpha\beta} + (\p_{\mu}\p_{\nu}-\eta_{\mu\nu}\p^2) H-M^2 (H_{\mu\nu}-H\eta_{\mu\nu})=0 
\label{qgadd1}
\end{equation}
where $H= \eta^{\mu\nu}H_{\mu\nu}$ above. Acting $\p^{\mu}$ from the left, we find the constraint
\be
\p^{\nu}\left(H_{\mu\nu}- \eta_{\mu\nu}H\right)=0 
\label{qgadd2}
\ee
Plugging this back into Eq.~\ref{qgadd1}, we have 
\be
\p^2H_{\mu\nu} - \p_{\mu}\p_{\nu}H -M^2\left(H_{\mu\nu}- \eta_{\mu\nu}H\right)=0,
\label{qgadd3}
\ee
which yields another constraint $H=H_{\mu\nu}\eta^{\mu\nu}=0$. Thus, the two constraints, $\p_{\mu}H^{\mu\nu}=0=H$, implies that $H_{\mu\nu}$ has five independent degrees of freedom or polarisations in four spacetime dimensions. Denoting the corresponding symmetric polarisation tensors by $\e_{\mu\nu}(k)$, 
these constraints imply that $k_{\mu}\e^{\mu\nu}(k)=k_{\nu}\e^{\mu\nu}(k)=0$, and $\e^{\mu}{}_{\mu}(k)=0$. We will come to the matter of explicit representation of these polarisation tensors in \ref{S20a}. 
 \\

\noindent
We will take the propagator for the massive spin-2 field as~\cite{vanDam:1970vg},
\begin{equation}
    \Delta^{M}_{\mu\nu\alpha\beta}= -\frac{i {\cal P}^{M}_{\mu\nu\alpha\beta}}{k^2+M^2}, 
\label{qg12'}
\end{equation}
where
\begin{equation}
    {\cal P}^{M}_{\mu\nu\alpha\beta}:= \frac12\left(\eta_{\mu\alpha}\eta_{\nu\beta}+\eta_{\mu\beta}\eta_{\nu\alpha}-\frac23\eta_{\mu\nu}\eta_{\alpha\beta}\right)
    \label{PM}
\end{equation}
Comparing the above propagator for the massive spin-2 field with that of the graviton  (Eq.~\ref{qg4}), we see that there is no smooth massless limit for the first. This leads to the famous vDVZ discontinuity in the bending of light~\cite{vanDam:1970vg, Zakharov}.

We will compute the amplitudes for the 2-2 scattering processes between  a massive spin-2 field and chiefly with a massive scalar. The incoming(outgoing) momenta will be denoted by $k_1, k_2\ (k'_1, k'_2)$, where $k_1, \ k'_1$ are the momenta of the massive spin-2 field. We have 
\begin{equation*}
k_1+k_2= k'_1+k'_2
\end{equation*}
We will denote the transfer momentum, $k_1-k_1'=k_2'-k_2$, carried by the internal graviton lines by $q$. In the non-relativistic limit, we have $q^0 \approx 0$. The gravitational potential, as per the first Born approximation, is defined via the Fourier transform of the non-relativistic limit of the 2-2  scattering Feynman amplitude~\cite{Bjerrum-Bohr:2002gqz},
\begin{equation}
\label{potential f}
V(\vec{r})=-\frac{1}{2m_{1}}\frac{1}{2m_{2}} \int \frac{d^3 \vec{q}}{(2\pi)^3} e^{-i\vec{q}\cdot \vec{r}} {\cal M}(\kappa, \vec{q}, m_{1},m_{2})\vert_{\rm NR}. 
\end{equation}
We will also require the following Fourier transforms,
\begin{equation}
\label{FT}
    \int\dfrac{d^3 \vec{q}}{(2 \pi)^3}\dfrac{ e^{i \vec{q}\cdot \vec{r}} }{|\vec{q}|} = \dfrac{1}{2 \pi^2 r^2}; \qquad 
    \int\dfrac{d^3 \vec{q}}{(2 \pi)^3}\dfrac{ e^{i \vec{q}\cdot \vec{r}} }{\vec{q}^2} = \dfrac{1}{4 \pi r}; \qquad 
    \int\dfrac{d^3 \vec{q}}{(2 \pi)^3} e^{i \vec{q}\cdot \vec{r}}\ln \vec{q}^2 = - \dfrac{1}{2 \pi r^3}.
\end{equation}
The various interaction terms between matter fields and the graviton have been listed in \ref{A0}. With these ingredients, we are now ready to go into the scattering calculations. We first wish to compute the tree level scattering amplitude of the massive spin-2 field with massive spin-0, spin-1, and spin-1/2 fields.

\section{Scattering at tree level}\label{S3}
\begin{figure}[h]
    \centering
    \includegraphics[width=0.35\linewidth]{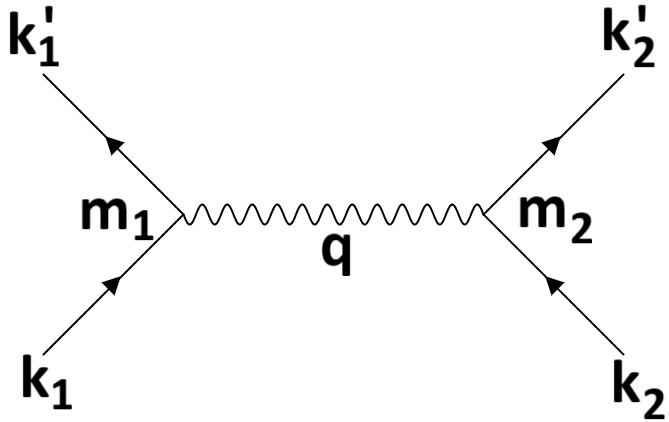}
    \caption{\small \it The tree level diagram of massive spin-0-spin-0 scattering. The straight lines stand for the scalar field, whereas the wavy line represents the graviton.  $q= k_1-k'_1=k_2'-k_2$ is the transfer momentum.}
    \label{fig0}
\end{figure}

In order to clarify notations, let us first review the well known scattering of two massive scalars, \ref{fig0}, and derive the Newton potential. 
Using Eq.~\ref{qg4}, and for the two scalar one graviton vertex, e.g.~\cite{Bjerrum-Bohr:2002gqz}, 
\begin{equation}
V_{\text{spin-0}}^{(1)\,\mu \nu} (k,k',m)= -\frac{i\kappa}{2} \Big[k^{\mu}k'^{\nu}+k'^{\mu}k^{\nu}-\eta^{\mu\nu}\left(k\cdot k' +m^2 \right) \Big],
\label{qg5'}
\end{equation}
The Feynman amplitude for this process is given by 
\begin{equation}
    \begin{split}
        i {\cal M}^{(1)}(\kappa^2, q, m_1,m_2)& =   V_{\text{\:spin-0}}^{(1)\,\mu \nu} (k_1,k_1',m_1)\Delta_{\mu\nu\alpha\beta}(q) V_{\text{\:spin-0}}^{(1)\,\alpha \beta} (k_2,k_2',m_2)\\
        &=-\frac{i\kappa^2}{8 q^2} \Big[k_1^{\mu}k_1'^{\nu}+k_1'^{\mu}k_1^{\nu}-\eta^{\mu\nu}\Big(k_1\cdot k_1' + m_1^2\Big) \Big] \Big( \eta_{\mu\alpha}\eta_{\nu\beta}+\eta_{\mu\beta}\eta_{\nu\alpha}-\eta_{\mu\nu}\eta_{\alpha\beta}\Big)\Big[k_2^{\alpha}k_2'^{\beta}\\
        & +k_2'^{\alpha}k_2^{\beta}-\eta^{\alpha\beta}\Big(k_2\cdot k_2' + m_2^2 \Big) \Big]\\
        &=-\frac{i\kappa^2}{2q^2}\Big[(k_1\cdot k_2) (k_1'\cdot k_2')+(k_1\cdot k_2') (k'_{1}\cdot k_2)-k_1\cdot k'_{1} \Big(k_2\cdot k_2' -m_2^2 \Big)  +k_2\cdot k_2'  \\
        & \Big((d/2-2)k_1\cdot k_1' -(d/2-1)m_1^2 \Big)-\frac{d}{2}\Big(k_2\cdot k_2' -m_2^2 \Big)\Big((d/2-2)k_1\cdot k_1'-(d/2-1)m_1^2 \Big) \Big]
\label{qg6}
    \end{split}
\end{equation}
In the non-relativistic limit, we have $k_1\cdot k'_1 \approx  -m^2_{1}$, $k_2\cdot k'_2 \approx -m^2_{2}$, $k_1\cdot k_2 \approx -m_1 m_2$ etc. and $ q^{\mu} \approx \vec{q}$, and we obtain
\begin{equation}
i {\cal M}^{(1)}(\kappa^2, q, m_1,m_2)\vert_{\rm NR} \approx   \frac{i\kappa^2 m_1^2 m_2^2}{2\vec{q}^2}
\label{qg7}
\end{equation}
The potential can be found using Eqs.~\ref{potential f}, \ref{FT},
\begin{equation}
V(r)= -\frac{\kappa^2 m_1m_2}{8} \int \frac{d^3 \vec{q}}{(2\pi)^3} \frac{e^{-i\vec{q} \cdot \vec{r}}}{\vec{q}^2}=-\frac{Gm_1m_2}{r}
\label{qg8}
\end{equation}
We next wish to compute the tree level scattering of a massive spin-2 field with massive spin-0, spin-1, and spin-1/2 fields.
\subsection{Massive spin-2 and spin-0 scattering}\label{S20a}
%
\begin{figure}[h]
    \centering
    \includegraphics[width=0.35\linewidth]{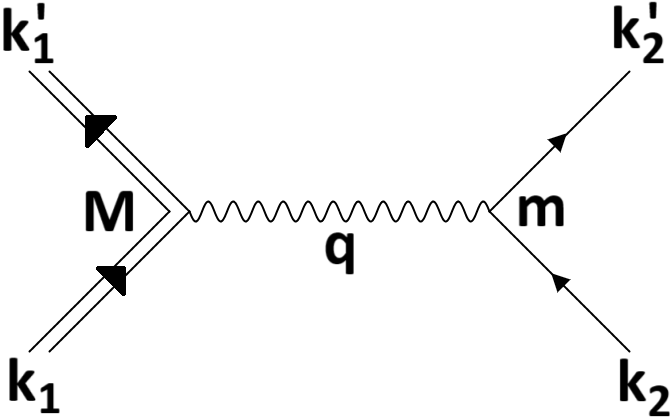}
    \caption{\small \it The tree level scattering  of massive spin-2-spin-0 fields. The double lines represent the massive spin-2 field.}
    \label{fig1}
\end{figure}
\noindent
Let us now compute the lowest order scattering amplitude between a massive scalar and a massive spin-2 field, \ref{fig1}.  This requires cubic vertex functions for the two scalar-one graviton, Eq.~\ref{qg5'}, and two massive spin-2-one graviton interactions. The first contains two indices, whereas it is clear that the second will contain six indices. A pair of them will correspond to the graviton, whereas the remaining four will pairwise correspond to the massive spin-2 field. However, note that these latter indices must be contracted with the polarisation of the massive spin-2 field. Due to this reason, and in order to make things look simple, we would retain those polarisation tensors in the three point vertex function (also in the four point vertex later), and {\it effectively} would represent the same with two indices only, and will denote it as $V^{(1)\,\text{spin-2}}_{\mu \nu,{\rm eff.}}(k,k',M)$.   It   can be read off from Eq.~\ref{A02},
 \begin{equation}
    \begin{split}
        V^{(1)\text{spin-2}}_{\mu ' \alpha ',{\rm eff.}}(k,k',M) = & - \frac{i\kappa}{2}\Big[2k \cdot k' \e_{(\mu'}{}^{\nu}(k )\e^{\star}_{\alpha') \nu}(k' ) - (k_{(\mu'} k'_{\alpha')}- \eta_{\mu'\alpha'}k \cdot k' ) \e_{\alpha\beta}(k )\e^{\star \alpha\beta}(k')  +M^2 (2 \e^{\mu}{}_{(\mu'}(k )  \\
        & \times \e^{\star}_{\alpha')\mu}(k' ) +\eta_{\mu'\alpha'} \e^{\alpha\beta}(k )\e^{\star}_{\alpha\beta}(k' )) + q^2\e_{\nu(\alpha'}(k )\e^{\star \nu}{}_{\mu')}(k' ) + \Big(k_{(\mu'}\eta_{\alpha')\alpha}  +k'_{(\mu'}\eta_{\alpha')\alpha} \Big) \\\
        & \Big(\e^{\alpha}{}_{\nu}(k )q_{\mu}\e^{\star \mu\nu}(k' ) - q_{\mu}\e^{\mu}{}_{\nu}(k )\e^{\star \alpha\nu}(k' ) \Big) \Big]
    \end{split}
\end{equation}
Using the above expression, Eqs.~\ref{qg4}, \ref{qg5'}, and also noting that the external momenta are on-shell, we write down the Feynman amplitude in the leading order  as, 
\begin{equation*}
    \begin{split}
        i {\cal M}^{(1)}(\kappa^2, q,M, m) =& V^{(1)\,\text{spin-2}}_{\mu ' \alpha',{\rm eff.}}(k_1,k_1',M) \Delta^{\mu ' \alpha '\nu' \beta'} (q) V_{\nu' \beta'}^{(1)\,\text{spin-0}} (k_2,k_2',m)\\
        & = -\frac{i\kappa^2}{8}\frac{(\eta^{\mu'\nu'}\eta^{\alpha'\beta'}+ \eta^{\mu'\beta'}\eta^{\nu'\alpha'}- \eta^{\mu'\alpha'}\eta^{\nu'\beta'})(k_{2\nu'}k'_{2\beta'}+k'_{2\nu'}k_{2 \beta'}- \eta_{\nu\beta'}(k_2\cdot k'_2+m^2) )}{q^2}
    \end{split}
\end{equation*}
\begin{eqnarray}
    \begin{split}
        &\times \Big[2k_1\cdot k'_1\e_{(\mu'}{}^{\nu}(k_1)\e^{\star}_{\alpha') \nu}(k'_1) - (k_{1(\mu'} k'_{1\alpha')}- \eta_{\mu'\alpha'}k_1\cdot k'_1) \e_{\alpha\beta}(k_1)\e^{\star \alpha\beta}(k'_1) +M^2 (2\e^{\mu}{}_{(\mu'}(k_1)\e^{\star}_{\alpha')\mu}(k'_1)  +\eta_{\mu'\alpha'} \e^{\alpha\beta}(k_1) \\
        & \times  \e^{\star}_{\alpha\beta}(k'_1)) + q^2\e_{\nu(\alpha'}(k_1)\e^{\star \nu}{}_{\mu')}(k'_1) + \Big\{k_{1(\mu'}\eta_{\alpha')\alpha}+k'_{1(\mu'}\eta_{\alpha')\alpha} \Big\}  \Big\{\e^{\alpha}{}_{\nu}(k_1)q_{\mu}\e^{\star \mu\nu}(k'_1) - q_{\mu}\e^{\mu}{}_{\nu}(k_1)\e^{\star \alpha\nu}(k'_1) \Big\} \Big]
\label{qgs1}
    \end{split}
\end{eqnarray}
Before we proceed, let us clarify the issue of the polarisation tensors $\e_{\mu\nu}(k)$, appearing above. 
In the {\it  rest frame}, the  five independent polarisation tensors  can be represented  by five trace free symmetric  matrices as~\cite{vanDam:1970vg}, 
\begin{eqnarray}
    \begin{split}
        \epsilon_{\mu \nu}^{(1)} = \sqrt{\frac{2}{3}}\begin{pmatrix}
            0 & 0 & 0 & 0 \\
            0 & \dfrac{1}{2} & 0 & 0  \\
            0 & 0 & \dfrac{1}{2} & 0 \\
            0 & 0 & 0  & -1 
        \end{pmatrix}, \hspace{2cm} \epsilon_{\mu \nu}^{(2)} =\frac{1}{\sqrt{2}}\begin{pmatrix}
            0 & 0 & 0 & 0 \\
            0 & 1 & 0 & 0 \\
            0 & 0 & -1 & 0 \\
            0 & 0  & 0 & 0
        \end{pmatrix}, \hspace{2cm} \epsilon_{\mu \nu}^{(3)} = \frac{1}{\sqrt{2}}\begin{pmatrix}
            0 & 0 & 0 & 0 \\
            0 & 0 & 1 & 0 \\
            0 & 1 & 0 & 0 \\
            0 & 0  & 0 & 0
        \end{pmatrix},  \\
        \epsilon_{\mu \nu}^{(4)} = \frac{1}{\sqrt{2}}\begin{pmatrix}
            0 & 0 & 0 & 0 \\
            0 & 0 & 0 & 1 \\
            0 & 0 & 0 & 0 \\
            0 & 1 & 0 & 0
        \end{pmatrix},  \hspace{2cm} \epsilon_{\mu \nu}^{(5)} =\frac{1}{\sqrt{2}}\begin{pmatrix}
            0 & 0 & 0 & 0 \\
            0 & 0 & 0 & 0 \\
            0 & 0 & 0 & 1 \\
            0 & 0 & 1 & 0
        \end{pmatrix}. \hspace{3.5cm}
    \end{split}
     \label{polaten}
\end{eqnarray}
The inner product between different polarisations  is defined via the trace norm as 
\begin{equation}
       \e^{(a)}_{\mu \nu}\e^{\star \mu\nu}_{(a')} \equiv {\rm  Tr }\left( \e_a^{\dagger} \e_{a'}\right)=    \delta_{aa'}\qquad (a,a'=1,2,3,4,5).
       \label{ip}
\end{equation}
For any momentum $\vec{k}$, the corresponding polarisation tensors can be found by boosting them through appropriate Lorentz transformation matrix, $\e_{\mu\nu}\to \Lambda_{\mu}{}^{\alpha}(\vec{k})\Lambda_{\nu}{}^{\beta}(\vec{k})\e_{\alpha\beta}$. Clearly, Eq.~\ref{ip} remains invariant under this boost transformation,
$$\e^{(a)}_{\mu \nu}(\vec{k})\e^{\star \mu\nu}_{(a')}(\vec{k}) = \delta_{aa'}$$
Note also that the rest frame representation Eq.~\ref{polaten} trivially satisfies $k^{\mu}\e_{\mu\nu}=0=k^{\nu}\e_{\mu\nu}$. These tensor equations will hold in the boosted frame as well.

Let us now try to understand the non-relativistic limit of the boosted polarisation tensors. Writing $\Lambda_{\mu}{}^{\alpha}= \delta_{\mu}{}^{\alpha}+\omega_{\mu}{}^{\alpha}$, for a small Lorentz boost, we have from  Eq.~\ref{polaten},
$$\e_{00}(\vec{k})=0, \qquad \e_{0i}(\vec{k})=\omega_{0j}(\vec{k}) \e_{ji}.$$
Since $\e_{i0}(\vec{k}=0)=0$,   in the product of two polarisation tensors, we may safely keep the pure spacelike components of $\e_{\mu\nu}(\vec{k})$ in the non-relativistic limit up to ${\cal O}(\omega)$.\\

\noindent
We next simplify the scattering amplitude of Eq.~\ref{qgs1} as
\begin{eqnarray}
 i {\cal M}^{(1)}_{aa'}=\frac{16 i \pi G }{q^2}\Big[m^2 M^2\e^{(a)}_{\mu \nu}(k_1)\e_{(a')}^{\star \mu\nu}(k'_1) + 2m (m+4M)q_{\mu}q_{\nu} \e^{(a)\, \mu}_{ \, \alpha}(k_1)\e^{\star \nu \alpha}_{(a')}(k'_1) + 8 m M \Big(k_{2\alpha}q_{\mu} -k'_{2\mu} q_{\alpha}\Big)\e^{(a)\, \alpha}_{ \, \nu}(k_1)\e^{\star \mu\nu}_{(a')}(k'_1) \Big].\nonumber\\
\end{eqnarray}
Note that the representation of Eq.~\ref{polaten} is real. Thus the complex conjugation symbol appearing above is spurious. However, we wish to keep them just to remind us that a particle with momentum $k'_1$ is created here, whereas the one without complex conjugation indicates annihilation of a particle with momentum $k_1$.
Note that the leading contribution to the potential comes from the first term on the right hand side of the above equation. Taking the non-relativistic as well as {\it rest frame} limit, we have 
\begin{equation}
    i {\cal M}^{(1)}(\kappa^2, q,M, m)\vert_{\rm NR} = \frac{i\kappa^2m^2 M^2}{2\vec{q}^2}  \delta_{aa'},
    \label{lead2-0}
\end{equation}
which reproduces the Newton potential.

The general form of  the long range gravitational potential reads
\begin{eqnarray}
    \begin{split}
        V = -    \frac{G m M}{ r}  \e^{(a)}_{i j}(\vec{k}_1)\e_{(a')}^{\star ij}(\vec{k'}_1)
        - \frac{2  G  m}{M r^3 }   \bigg(\delta_{ij} -  \frac{3 r^{i} r^{j}}{r^2}  \bigg)  \e^{(a)\, i}_{ k}(\vec{k}_1)\e^{\star j k}_{(a')}(\vec{k}'_1) + {\cal O}(r^{-4}). 
    \end{split}
\end{eqnarray}
%

\subsection{Massive spin-2 and massive spin-1 scattering}\label{S21}

\noindent
From Eq.~\ref{A02}, we read off the one graviton two massive spin-1 vertex~\cite{Holstein:2008sx}, 
\begin{equation}
    \begin{split}
         V_{\mu \nu}^{(1)\,\text{spin-1}}(k,k',m_v) = \frac{i\kappa}{2} \left[- k \cdot \e'\ \e_{(\mu'}k'_{\nu)}- k' \cdot \e \ \e'_{(\mu}k_{\nu)} + \e\cdot \e'\ k_{(\mu}k'_{\nu)} +\eta_{\mu\nu}k \cdot \e' \ k' \cdot \e\right],
    \end{split}
\end{equation}
where we have used the fact that the external momenta are on-shell and non-relativistic, and $\e_{\mu}$, $\e^{\star}_{\mu}$'s are the polarisation vectors for the massive spin-1 field. 
Remembering the constraint $k\cdot \e=0= k'\cdot \e'$, the Feynman amplitude for this process is written as,
\begin{eqnarray}
    \begin{split}
        i {\cal M}^{(1)}(\kappa^2, q,& M, m_{v}) = V^{(1)\,\text{spin-2}}_{\mu \alpha,{\rm eff.}}(k_1,k_1',M) \Delta^{\mu \alpha\nu \beta} (q) V_{\nu \beta}^{(1)\,\text{spin-1}}(k_2,k_2',m_v)\\
        =&-\frac{i\kappa^2(\eta^{\mu'\nu'}\eta^{\alpha'\beta'}+ \eta^{\mu'\beta'}\eta^{\nu'\alpha'}- \eta^{\mu'\alpha'}\eta^{\nu'\beta'})}{8q^2}
        \Big[-k_{1(\mu'}k'_{1\alpha')} \e_{\mu\nu}(k_1)\e^{\star \mu\nu}(k'_1) +(k_1+k'_1)_{(\mu'}\eta_{\alpha')\alpha}\\ 
        & \times \Big\{\e^{\alpha}{}_{\nu}(k_1)q_{\mu}\e^{\star \mu\nu}(k'_1)- q_{\mu}\e^{\mu}{}_{\nu}(k_1)\e^{\star \alpha\nu}(k'_1) \Big\} \Big] 
        \Big[- k_2 \cdot \e'_2\ \e_{(\nu'}(k_2)k'_{2\beta')}- k'_2 \cdot \e_2 \ \e^{\star}_{(\nu'}(k'_2)k_{2\beta')}\\ 
        &  + \e_2\cdot \e'_2\ k_{2(\nu'}k'_{2\beta')} +\eta_{\nu'\beta'}k_2 \cdot \e'_2 \ k'_2 \cdot \e_2 \Big]\\
        =&  {\frac{i G \pi}{q^2}\Bigg[  16  M^2 m_v^2 \vec{\epsilon_2 }.\vec{\epsilon_2}' - 8   M^2   i (\vec{k}\times \vec{q}).\vec{S} + 4  M^2    \vec{q}.\vec{\epsilon_2 }' \vec{q}.\vec{\epsilon_2 } 
        \bigg] \epsilon^{(a)}_{ij} (k_2)\epsilon ^{\star ij }_{(a')}}(k'_2) \ + \ {\rm sub-leading ~ contributions}.
    \end{split}
\end{eqnarray}
Since $\vec{k}_1- \vec{k}'_1= \vec{k}'_2-\vec{k}_2=\vec{q}$,  we have written above
\begin{equation}
\vec{k}_1= \vec{k} + \frac{\vec q}{2}, \qquad \vec{k}'_1= \vec{k} - \frac{\vec q}{2}; \qquad \qquad \vec{k}_2= \vec{k} - \frac{\vec q}{2}, \qquad \vec{k}'_2= \vec{k} + \frac{\vec q}{2},
\label{qgs2}
\end{equation}
where $\vec{k}$ should be interpreted as the momentum of the centre of momentum. We also write  in the non-relativistic limit for the polarisation vectors of the massive spin-1 field following~\cite{Holstein:2008sx},
\be
 \e^{\mu}(k)\vert_{\rm NR} \simeq \left( \frac{\vec{k}\cdot \vec{\e}}{m_v},\ \vec{\e}\right) 
 \label{pol}
 \ee
which trivially satisfies, $k\cdot \e=0$ in the non-relativistic limit, $k^0 \approx -M$. We have
\begin{equation}
\epsilon_{\mu}(k_1)\,\epsilon^{\,\star \, \mu}(k'_1)\vert_{\rm NR}= \vec{\epsilon}\,(\vec{k}_1)\cdot \vec{\epsilon}^{\,\,\star}(\vec{k}'_1)- \frac{i}{2m_v^2}\,\vec{S} \cdot (\vec{k}\times \vec{q})-\frac{1}{m_v^2} \, \vec{k}\cdot \vec{\epsilon}\,(\vec{k}_1)\ \vec{k}\cdot \vec{\epsilon}^{\,\,\star}(\vec{k}'_1) +\frac{1}{4m_v^2} \, \vec{q}\cdot \vec{\epsilon}\,(\vec{k}_1)  \ \vec{q}\cdot \vec{\epsilon}^{\,\,\star}(\vec{k}'_1),
\label{qgs3}
\end{equation}
where we have written
\be
-i\vec{S} = \vec{\epsilon}\,(\vec{k}_1)\times \vec{\epsilon}^{\,\,\star}(\vec{k}'_1),
\label{qgs3'}
\ee
and have  used the trivial identity,
$$\vec{q}\cdot \vec{\e} \ \vec{k}\cdot \vec{\e}'^{\star} = \vec{k}\cdot \vec{\e} \ \vec{q}\cdot \vec{\e}'^{\star}+ i\vec{S}\cdot (\vec{k}\times \vec{q})$$
Putting everything together now, we have the two body long range gravitational potential,
\begin{eqnarray}
    \begin{split}
         V(r)\vert_{\rm Tree} =& \bigg[  -\frac{G m_v M  }{r} 
         \vec{\epsilon_2 }.\vec{\epsilon_2}'-   \frac{G M   }{2 m_v r^2} 
         (\vec{k}\times \hat{r}).\vec{S}
        +  \frac{3 G M   }{4 m_v r^5} 
        \vec{r}.\vec{\epsilon_2 } \vec{r}.\vec{\epsilon}'_2   \bigg] \epsilon^{(a)} _{i j }(\vec{k}_1) \epsilon^{\star i j }_{(a')}(\vec{k}'_1)\ + \ {\cal O}(r^{-4}).  
    \end{split}
\end{eqnarray}
%
\subsection{Massive spin-2 and spin-1/2 scattering}\label{S2.5}
Let us now consider the massive spin-1/2 and massive spin-2 scattering at the lowest order. We have the one graviton two fermion vertex in the momentum space~\cite{Holstein:2008sx},
\begin{equation}
V_{\text{spin-1/2}}^{(1)\,\mu\nu}(k,k',m_f)= -\frac{i\kappa}{2} \left[ \frac14\gamma^{(\mu}(k+k')^{\nu)}- \eta^{\mu\nu} \left(\frac{\slashed{k}+\slashed{k'}}{2}-m_f \right)  \right].
\label{qg26}
\end{equation}
The tree level scattering amplitude reads,
\begin{eqnarray}
    \begin{split}
        i {\cal M}^{(1)}(\kappa^2, q,M, m_{f})=& V^{(1)\,\text{spin-2}}_{\mu ' \alpha',{\rm eff.}}(k_1,k_1',M) \Delta^{\mu ' \alpha '\nu' \beta'} (q) V_{\nu' \beta'}^{(1)\,\text{spin-1/2}}(k_2,k_2',m_f)\\
        &=-\frac{i\kappa^2(\eta^{\mu'\nu'}\eta^{\alpha'\beta'}+ \eta^{\mu'\beta'}\eta^{\nu'\alpha'}- \eta^{\mu'\alpha'}\eta^{\nu'\beta'})}{8q^2}  \Big[2k_1\cdot k'_1\e_{(\mu'}{}^{\nu}(k_1)\e^{\star}_{\alpha') \nu}(k'_1) - (k_{1(\mu'} k'_{1\alpha')} \\
        &  - \eta_{\mu'\alpha'}k_1\cdot k'_1) \e_{\alpha\beta}(k_1)\e^{\star \alpha\beta}(k'_1) +M^2 (2\e^{\mu}{}_{(\mu'}(k_1)\e^{\star}_{\alpha')\mu}(k'_1)  +\eta_{\mu'\alpha'} \e^{\alpha\beta}(k_1)  \e^{\star}_{\alpha\beta}(k'_1)) \\
        &   + q^2\e_{\nu(\alpha'}(k_1)\e^{\star \nu}{}_{\mu')}(k'_1) + \Big\{k_{1(\mu'}\eta_{\alpha')\alpha}+k'_{1(\mu'}\eta_{\alpha')\alpha} \Big\}  \Big\{\e^{\alpha}{}_{\nu}(k_1)q_{\mu}\e^{\star \mu\nu}(k'_1) - q_{\mu}\e^{\mu}{}_{\nu}(k_1)\e^{\star \alpha\nu}(k'_1) \Big\} \Big] \\
        & \times \bar{u}_{s'}(k'_2) \Big[ \frac14\gamma_{(\nu'}(k_2+k'_2)_{\beta')}- \eta_{\nu'\beta'}\Big(\frac{\slashed{k_2}+\slashed{k'_2}}{2}-m_f\Big) \Big] u_s(k_2)\\
        =&  \frac{ i  G \pi}{q^2} \bigg[ \bigg( 16    m^2 M^2  \epsilon _{\mu \nu }^{(a)} \epsilon ^{\star \mu \nu }_{(a')}   -112    m^2 k_{\mu } q_{\alpha }  \epsilon _{\nu }^{(a)\,\alpha } \epsilon ^{\star \mu \nu }_{(a')}-16    m^2 k_{\mu } q_{\alpha }  \epsilon _{\nu }^{(a)\, \mu } \epsilon ^{\star \nu \alpha }_{(a')} +128    m^2 q_{\mu } \epsilon _{\nu \alpha }^{(a)} k^{\nu }  \epsilon ^{\star \mu \alpha }_{(a')}  \\
        &   -16    m M q_{\alpha } k^{\mu }  \epsilon _{\nu }^{(a)\, \alpha } \epsilon _{(a')\, \mu }^{\star \nu }+16    m M q_{\alpha } k^{\mu }  \epsilon _{\mu \nu }^{(a)} \epsilon ^{\star \nu \alpha }_{(a')}  \bigg) \delta _{ss'}
        +  \bigg( \frac{8     M q_{\alpha } k^{\mu } \epsilon _{\nu }^{(a)\, \alpha } \epsilon _{(a')\, \mu }^{\star \nu } }{m }-\frac{8     M q_{\alpha } k^{\mu } \epsilon _{\mu \nu }^{(a)} \epsilon ^{\star \nu \alpha }_{(a')} }{m } \\
        &   +8     k_{\mu } q_{\alpha } \epsilon _{\nu }^{(a)\, \alpha } \epsilon ^{\star \mu \nu }_{(a')} +16     m M \epsilon _{\mu \nu }^{(a)} \epsilon ^{\star \mu \nu }_{(a')}  -12     M^2 \epsilon _{\mu \nu }^{(a)} \epsilon ^{\star \mu \nu }_{(a')} 
         -8     k_{\mu } q_{\alpha } \epsilon _{\nu }^{(a)\, \mu } \epsilon ^{\star \nu \alpha }_{(a')} \bigg) i\vec{k}\times \vec{q}.\vec{S}_{1/2} \bigg] 
\label{qgf1}
    \end{split}
\end{eqnarray}
where $s=\pm$, and we used the Gordon identity, 
\be
\bar{u}_{s'}(k'_2) \gamma^{\mu} u_s(k_2)=\frac{1}{2m_f}\bar{u}_{s'}(k'_2)\left[ (k_2+k'_2)^{\mu}-\frac12 [\gamma^{\mu},  \gamma^{\nu}]q_{\nu}\right]u_s(k_2).
\label{qgf2}
\ee
We also take.
\begin{equation}
u_{s}(\vec{k})=
\sqrt{E_{\vec k}+m_f}\begin{pmatrix}
1 \\
\\
\dfrac{\vec{k}\cdot \vec{\sigma}}{E_{\vec k}+m_f}
\end{pmatrix} \chi_{s}
\label{qgf3}
\end{equation}
where
\be
\chi_+=\begin{pmatrix}
1 \\
0
\end{pmatrix}, \qquad \chi_-= \begin{pmatrix}
0 \\
1
\end{pmatrix}.
\label{qgf4}
\ee
We next compute using Eq.~\ref{qgf3}, Eq.~\ref{qgf4}, in the non-relativistic limit,
\begin{equation}
\bar{u}_{s'}(\vec{k'}_2){u}_{s}(\vec{k}_2) =2m_f \left[\delta_{ss'} -\frac{i}{2m_f^2}(\vec{k}\times \vec{q})\cdot \vec{S}^{ss'}_{1/2}\right] + \ {\rm subleading~terms}
\label{qgf5}
\end{equation}
where we have defined the matrix element of the spin vector for the spin-1/2 field  as ($\hbar=1$),
$$
\vec{S}^{ss'}_{1/2}=\frac12 \chi^{\dagger}_{s'} \vec{\sigma}\chi_{s}.
$$
Putting things together now, the gravitational potential reads, 
\begin{eqnarray}
    \begin{split}
        V(r)\vert_{\rm Tree} = \left[ -\frac{G m_f M  }{r} \delta _{ss'} 
         + \Bigg( \frac{G  }{r^2}  -\frac{3 G M  }{4 m_f r^2}\Bigg) ( \vec{k}\times \hat{r}).\vec{S}\right]\epsilon^{(a)} _{i j } (\vec{k}_1)\epsilon ^{\star i j }_{(a')}(\vec{k}'_1) \ + \ {\cal O}(r^{-3}).
    \end{split}
\end{eqnarray}

\section{Massive spin-2 and spin-0 scattering at ${\cal O}(G^2)$}\label{S4}

We wish to compute below the one loop correction (${\cal O}(\kappa^4)$) to the 2-2 scattering for the massive spin-2 and spin-0 fields. The expressions are extremely lengthy compared to the tree level ones we have computed above, and we will present below only the explicitly  spin or polarisation  independent leading parts. The integrals necessary for the following computations are provided in \ref{B}.

\subsection{ The ladder and the cross-ladder diagrams}
\begin{figure}[h!]
\begin{center}
    \includegraphics[width=0.3\linewidth]{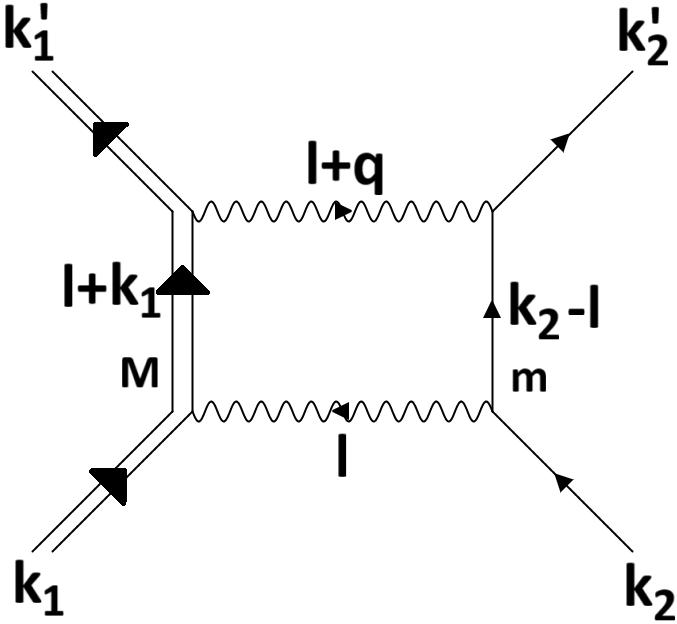} \hspace{3cm}
     \includegraphics[width=0.263\linewidth]{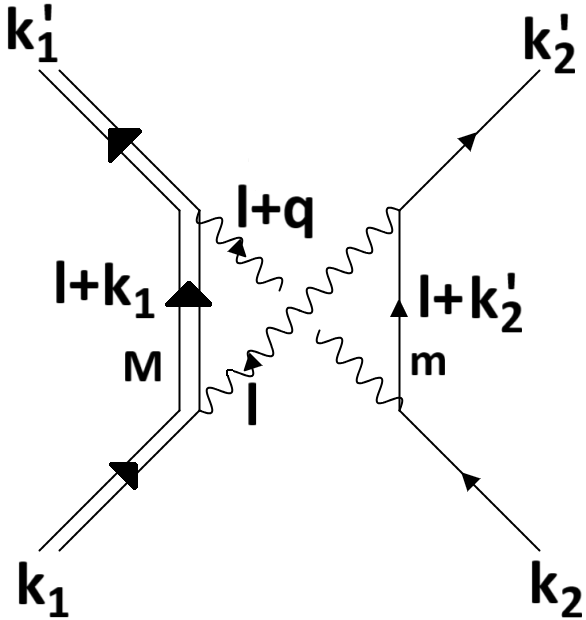} 
  \caption{\small \it The ladder (left) and the cross-ladder diagram for massive spin-2-spin-0 fields scattering. As earlier, double and single straight lines respectively stand for the massive spin-2 and spin-0 fields, whereas a wavy line represents the graviton.  $q= k_1-k'_1=k_2'-k_2$ is the transfer momentum.}
   \label{fl1}
\end{center}
\end{figure}

We begin by considering the ladder and cross-ladder diagrams, \ref{fl1}.  Note that a massive spin-2 field is created/annihilated with momentum $k_1' /k_1$ in this process. Accordingly, there will be $\e^{\star}, \e$ at these two vertices. The remaining two polarisations would sum up to yield the massive spin-2 propagator. 
Accordingly, the Feynman amplitudes for the ladder and the cross-ladder diagrams are respectively given by,  
\begin{equation}
    \begin{split}
        i \cal M_{\text{ladder}} =& \int\frac{d^4l}{(2\pi)^4}  \bar{V}^{(1)\,\text{spin-2}}_{\mu  \nu,{\rm eff.}}(k_1,l+k_1, M) \epsilon^\star _{\lambda' \eta '} (k'_1)\frac{-i {\cal P}_{M}^{\lambda' \eta ' \epsilon ' \psi' }}{(l+k_1)^2+M^2} \bar{V}^{(1)\,\text{spin-2}}_{\alpha \beta,{\rm eff.}}(l+k_1,k'_1, M) \epsilon _{\epsilon ' \psi'}(k_1)   \\
        & \times  V^{(1)\text{spin-0}}_{\gamma\delta}(k_2,k_2-l,m) V^{(1)\text{spin-0}}_{ \rho \sigma}(k_2-l,k'_2,m) \frac{-i}{(l-k_2)^2 + m^2}\frac{-i{\cal P}^{\mu\nu\gamma\delta}}{l^2}\frac{-i{\cal P}^{\alpha \beta \rho \sigma}}{(l+q)^2},
    \end{split}
\end{equation}
and
\begin{eqnarray}
    \begin{split}
        i \mathcal{M}_{\text{cross-ladder}} =& \int\frac{d^4l}{(2\pi)^4} \bar{V}^{(1)\,\text{spin-2}}_{\mu  \nu}(k_1,l+k_1, M) \epsilon^\star _{\lambda' \eta '}(k'_1) \frac{-i {\cal P}_{M}^{\lambda' \eta' \epsilon' \psi' }}{(l+k_1)^2+M^2} \bar{V}^{(1)\,\text{spin-2}}_{\alpha \beta}(l+k_1,k'_1, M) \epsilon _{\epsilon ' \psi'} (k_1) \\
        & \times V^{(1)\text{spin-0}}_{\rho \sigma}(k_2,l+ k'_2,m) V^{(1)\text{spin-0}}_{\gamma\delta}(l+ k'_2,k'_2,m)\frac{-i}{(l+ k'_2)^2 + m^2}\frac{{-i\cal P}^{\mu\nu\gamma\delta}}{l^2}\frac{-i{\cal P}^{\alpha \beta \rho \sigma}}{(l+q)^2},
    \end{split}
\end{eqnarray}
where  ${\cal P}_{M}$ is given by Eq.~\ref{PM}, and $\bar{V}^{(1)\,\text{spin-2}}_{\mu  \nu,{\rm eff.}}$ contains {\it no} polarisation tensor. We have, after some algebra, 
\begin{equation}
\begin{split}
    \mathcal{M}_{\text{lad. + cr.-lad. tot.}} =& \left[-\frac{16 \pi ^2 G^2 m^3 M^2 }{3 q}+\frac{10 \pi ^2 G^2 m^2 M^3 }{3 q}-\frac{46}{3} G^2 m^2 M^2  \ln q^2
       \right] \epsilon_{\mu \nu}^{(a)} (k_1) \epsilon^{\star \,\mu \nu \, (a')}(k'_1)  + \\
    & + \ {\rm polarisation-dependent~subleading~terms}.
\end{split}
       \end{equation}
The corresponding correction to the Newton potential  becomes,
\begin{equation}
\begin{split}
    V_{\text{lad. + cr.-lad. tot.}}(r) =  V_{\text{ladder}}(r) +V_{\text{cross-ladder}}(r) =& \Big[  \frac{2 G^2 m^2 M }{3 r^2}-\frac{5 G^2 m M^2 }{12 r^2}-\frac{23 G^2 m M }{12 \pi  r^3}   \Big] \epsilon_{ij}^{(a)}(\vec{k}_1) \epsilon^{\star \, ij \, (a')}(\vec{k}'_1) \\
    & + \ {\rm subleading~terms}.
    \label{box}
\end{split}    
\end{equation}
%

\subsection{ The triangle diagrams}
\begin{figure}[h!]
\begin{center}
    \includegraphics[width=0.3\linewidth]{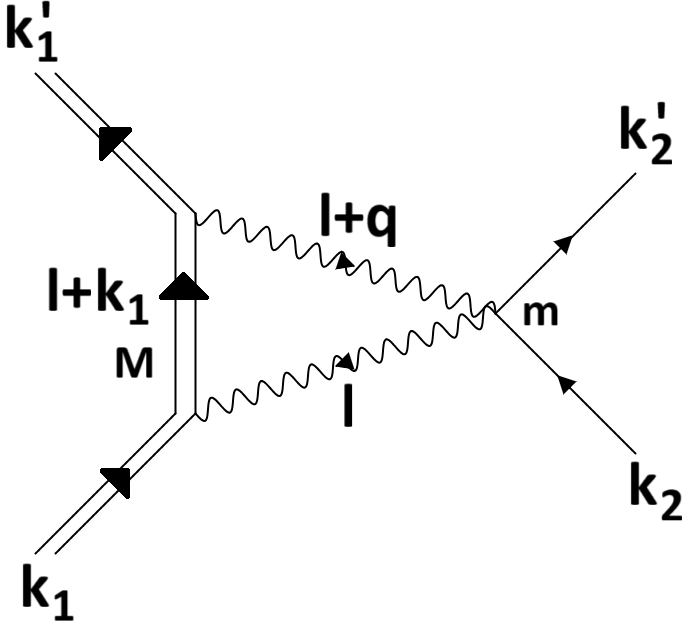} \hspace{3cm}
    \includegraphics[width=0.25\linewidth]{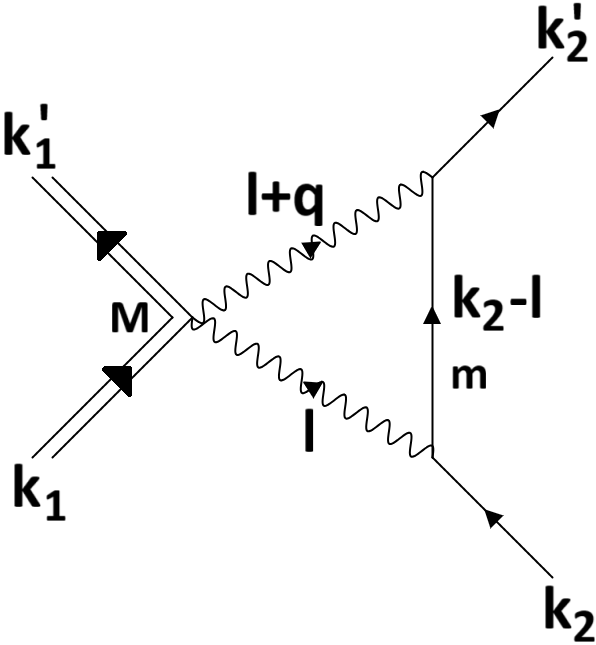} 
  \caption{\small \it The triangle diagrams  for massive spin-2-spin-0 fields scattering.   }
   \label{fl3}
\end{center}
\end{figure}

Let us next consider the triangle diagrams, \ref{fl3}. 
\noindent
We need to consider the $2$-scalar and $2$-graviton vertex for our purpose~\cite{Bjerrum-Bohr:2002gqz},
\begin{eqnarray}
    \begin{split}
    \label{qg21}
        V_{\text{\:spin-0}}^{(2)\,\mu \lambda \rho \sigma} (k,k',m) =& i\kappa^2 \Big[\Big\{ I^{\mu \lambda \alpha \nu}  I^{\rho\sigma \beta}{}_{\nu} -\frac14 \big(\eta^{\mu \lambda}I^{\rho\sigma \alpha \beta} +\eta^{\rho \sigma}I^{\mu \lambda \alpha \beta}  \big)  \Big\} \left(k_{\alpha}k'_{\beta}+k_{\beta}k'_{\alpha} \right)\\ 
        & -\frac12 \Big(I^{\mu \lambda \rho \sigma} -\frac12 \eta^{\mu \lambda}\eta^{\rho\sigma} \Big) \left(k\cdot k' +m^2 \right) \Big] 
    \end{split}
\end{eqnarray}
where,
$$I_{\mu\nu\lambda \rho}= \dfrac12 \left( \eta_{\mu\lambda} \eta_{\nu\rho} + \eta_{\mu\rho} \eta_{\nu\lambda}\right).$$ 
The Feynman amplitudes of triangle-1 and triangle-2 diagrams respectively reads, \ref{fl3}, 
\begin{equation}
    \begin{split}
        i\mathcal{M}_{\text{triangle-1}} =& \int\frac{d^4l}{(2\pi)^4} \bar{V}^{(1)\,\text{spin-2}}_{\mu  \nu, {\rm eff.}}(k_1,l+k_1, M) \epsilon^\star _{\lambda' \eta '}(k_1') \Bigg[-\frac{i {\cal P}_{M}^{\lambda' \eta ' \epsilon ' \psi' }}{(l+k_1)^2+M^2} \Bigg] \bar{V}^{(1)\,\text{spin-2}}_{\alpha \beta, {\rm eff.}}(l+k_1,k'_1, M) \epsilon _{\epsilon ' \psi'} (k_1)\\
        & \times V^{(2)\text{spin-0}}_{\sigma\rho\gamma\delta}(k_2,k'_2,m)\ \frac{-i{\cal P}^{\alpha\beta\gamma\delta}} {(l+q)^2}\ \frac{-i{\cal P}^{\mu\nu\sigma\rho}}{l^2}
    \end{split}
\end{equation}
and,
\begin{equation}
    \begin{split}
        i \mathcal{M}_{\text{triangle-2}} =& \int\frac{d^4l}{(2\pi)^4}  V_{(2)\text{spin-2}}^{\mu\nu\alpha\beta}(k_1,k'_1,M)  V_{(1)\text{spin-0}}^{\sigma\rho}(k_2,k_2-l,m)  V_{(1)\text{spin-0}}^{\gamma\delta}(k_2-l,k'_2,m) \\
        & \times \frac{ -i{\cal P}_{\mu\nu\sigma\rho}}{l^2}\ 
\frac{-i{\cal P}_{\alpha\beta\gamma\delta}}{(l+q)^2} \frac{-i}{(l-k_2)^2 + m^2},
    \end{split}
\end{equation}
where $V_{(2)\text{spin-2}}^{\mu\nu\alpha\beta}(k_1,k'_1,M)$ is the {\it effective} two massive spin-2-two graviton vertex containing two polarisations $\e_{\mu\nu}(k_1)$ and  $\e^{\star}_{\alpha\beta}(k'_1)$, and can be read off from Eq.~\ref{A03}. Note that the terms containing the derivative of the graviton field would yield terms containing the transfer momentum $q$, and such terms will make a subleading contribution to the potential.  Simplifying now the above equations, we obtain, 
\begin{equation}
\begin{split}
    \mathcal{M}_{\text{triangle-1}} =& \Big[-\frac{32 \pi ^2 G^2 m^3 M^2 }{3 q}  +  \frac{24 \pi ^2 G^2 m^2 M^3 }{q}  + 24 G^2 m^2 M^2 \ln q^2 \Big]  \epsilon_{\mu \nu}^{(a)}(k_1) \epsilon^{\star \,\mu \nu \, (a')}(k'_1)\\
    & + \ {\rm subleading~terms},
\end{split}
\end{equation}
and
\begin{equation}
\begin{split}
    \mathcal{M}_{\text{triangle-2}} =& \Big[\frac{3 G^2 m^3 M^2 \pi^2 }{ q}  + \frac{3G^2 m^2 M^3 \pi^2 }{q}  +3 G^2 m^2 M^2 \ln q^2  \Big]  \epsilon_{\mu \nu}^{(a)}(k_1) \epsilon^{\star \,\mu \nu \, (a')}(k'_1)\\
    & + \ {\rm subleading~terms}.
\end{split}
\end{equation}
The corresponding correction to the Newton potential in the non-relativistic limit becomes,
\begin{equation}
    \begin{split}
        V_{\text{triangle tot.}} (r) = V_{\text{triangle-1}} (r) + V_{\text{triangle-2}} (r) =& \Bigg[  \frac{27 G^2 m M }{8\pi  r^3}  + \frac{23 G^2 m^2 M  }{24 r^2}  - \frac{27 G^2 m M^2 }{8 r^2} \Bigg] \epsilon_{ij}^{(a)}(\vec{k}_1) \epsilon^{\star \, ij \, (a')}(\vec{k}'_1) \\
        & + \ {\rm subleading~terms}.
    \end{split}
\end{equation}
%

\subsection{The double seagull diagram}
\begin{figure}[h!]
\begin{center}
    \includegraphics[width=0.3\linewidth]{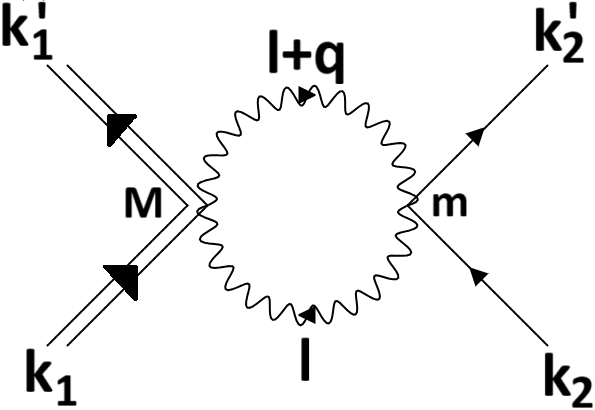} 
  \caption{\small \it The double seagull diagram for massive spin-2-spin-0 fields scattering.   }
   \label{fl5}
\end{center}
\end{figure}

We next consider the double seagull diagram of \ref{fl5}.
We have the Feynman amplitude,
\begin{equation}
    \begin{split}
       i \mathcal{M}_{\text{double-seagull}} =& {1\over 2!}\int{d^4l\over (2\pi)^4} V_{(2)\text{spin-2}}^{\alpha\beta\gamma\delta}(k_1,k'_1,M) V_{(2)\text{spin-0}}^{\sigma\rho\mu\nu}(k_2,k'_2,m) \frac{-i{\cal P}_{\alpha\beta\mu\nu}}{(l+q)^2} \frac{-i{\cal P}_{\gamma\delta\sigma\rho}}{l^2},
    \end{split}
\end{equation}
where as earlier $V_{(2)\text{spin-2}}^{\alpha\beta\gamma\delta}(k_1,k'_1,M)$ contains contractions with two polarisations of the massive spin-2 field. The above expression yields, 
\begin{equation}
        \mathcal{M}_{\text{double-seagull}} =-\frac{1}{8} G^2 m^2 M^2  \ln q^2  \epsilon_{\mu \nu}^{(a)} (k_1)\epsilon^{\star \,\mu \nu \, (a')}(k'_1) + \ {\rm subleading~terms}, 
\end{equation}
which gives the correction to the potential,
\begin{equation}
    V _{\text{double-seagull}} (r) = -\frac{G^2 m M}{64 \pi r^3} \epsilon_{ij}^{(a)}(\vec{k}_1)  \epsilon^{\star \, ij  \, (a')}(\vec{k}'_1) + \ {\rm subleading~terms}. 
\end{equation}
%

\subsection{The vertex correction diagram-1}
\begin{figure}[h!]
\begin{center}
    \includegraphics[width=0.3\linewidth]{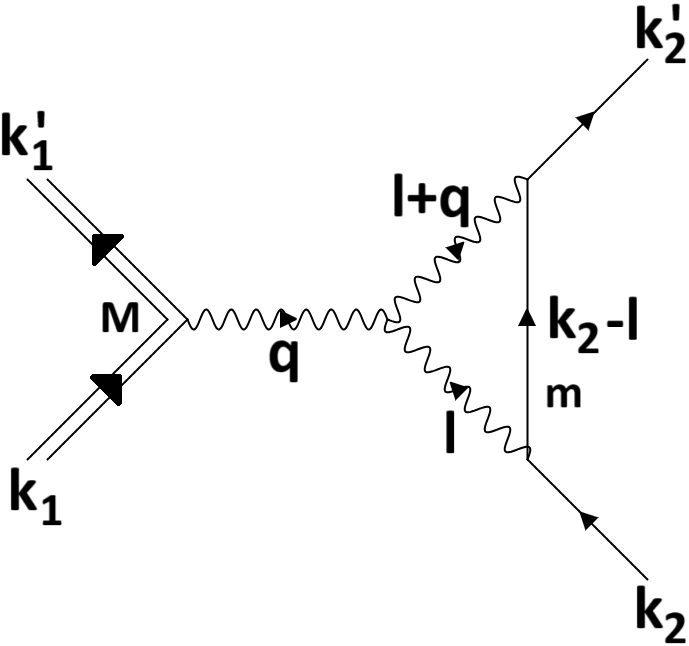} 
  \caption{\small \it The first vertex correction diagram for massive spin-2-spin-0 fields scattering.  }
   \label{fl6}
\end{center}
\end{figure}

Let us now come to the contribution from the vertex correction diagrams, the first of which is given by \ref{fl6}. To evaluate this diagram, we need to consider the three graviton vertex given by~\cite{Bjerrum-Bohr:2002gqz},
\begin{equation}
    \begin{split}
    \label{qg25}
        V^{(3)\mu\nu}_{\alpha\beta \gamma\delta}(k, q) =& -\frac{i\kappa}{2} \Big[  P_{\alpha\beta \gamma\delta}\Big(k^{\mu}k^{\nu}+(k-q)^{\mu}(k-q)^{\nu}+q^{\mu}q^{\nu} -\frac32 \eta^{\mu\nu}q^2\Big) +2 q_{\lambda} q_{\sigma} \Big(I_{\alpha\beta}{}^{\sigma\lambda}I_{\gamma\delta}{}^{\mu\nu}+I_{\gamma\delta}{}^{\sigma\lambda}I_{\alpha\beta}{}^{\mu\nu} \\
        & -I_{\alpha\beta}{}^{\mu\sigma}I_{\gamma\delta}{}^{\nu\lambda}- I_{\gamma\delta}{}^{\mu\sigma}I_{\alpha\beta}{}^{\nu\lambda}  \Big) +\Big\{q_{\lambda} q^{\mu}\Big(\eta_{\alpha\beta} I_{\gamma\delta}{}^{\nu\lambda}+\eta_{\gamma\delta} I_{\alpha\beta}{}^{\nu\lambda} \Big)   + q_{\lambda} q^{\nu}\Big(\eta_{\alpha\beta} I_{\gamma\delta}{}^{\mu\lambda}+\eta_{\gamma\delta} I_{\alpha\beta}{}^{\mu\lambda} \Big) \\
        & - q^2 \Big(\eta_{\alpha\beta} I_{\gamma\delta}{}^{\mu\nu}+\eta_{\gamma\delta} I_{\alpha\beta}{}^{\mu\nu}\Big) - \eta^{\mu\nu}q_{\sigma}q_{\lambda}\Big(\eta_{\alpha \beta}I_{\gamma\delta}{}^{\sigma\lambda}+ \eta_{\gamma\delta}I_{\alpha \beta}{}^{\sigma\lambda}\Big)  \Big\} + \Big\{-2q_{\lambda} \Big(I_{\alpha\beta}{}^{\lambda \sigma} I_{\gamma\delta \sigma}{}^{\nu} (k-q)^{\mu} \\
        & +I_{\alpha\beta}{}^{\lambda \sigma} I_{\gamma\delta \sigma}{}^{\mu} (k-q)^{\nu} +  I_{\gamma\delta}{}^{\lambda \sigma}I_{\alpha\beta\sigma}{}^{\nu}k^{\mu}+ I_{\gamma\delta}{}^{\lambda \sigma}I_{\alpha\beta\sigma}{}^{\mu}k^{\nu} \Big) +q^2 \Big(I_{\alpha\beta\sigma}{}^{\mu}I_{\gamma\delta}{}^{\nu\sigma}+ I_{\gamma\delta\sigma}{}^{\mu}I_{\alpha\beta}{}^{\nu\sigma} \Big) \\
        & +\eta^{\mu\nu}q_{\sigma}q_{\lambda}\Big(I_{\alpha\beta}{}^{\lambda\rho}I_{\gamma\delta \rho}{}^{\sigma}+ I_{\gamma\delta}{}^{\lambda\rho}I_{\alpha\beta\rho}{}^{\sigma} \Big)  \Big\} + \Big\{\Big(k^2+(k-q)^2 \Big) \Big(I_{\alpha\beta}{}^{\mu\sigma}I_{\gamma\delta\sigma}{}^{\nu} + I_{\gamma\delta}{}^{\mu\sigma}I_{\alpha\beta\sigma}{}^{\nu} \\
        & -\frac12 \eta^{\mu\nu}{\cal P}_{\alpha\beta\gamma\delta}\Big)- \Big(I_{\gamma\delta}{}^{\mu\nu}\eta_{\alpha\beta}k^2+I_{\alpha\beta}{}^{\mu\nu}\eta_{\gamma\delta}(k-q)^2  \Big)  \Big\} \Big]
    \end{split}
\end{equation}
%
The corresponding Feynman amplitude reads,
\begin{equation}
        \begin{split}
            i \mathcal{M}_{\text{vertex-correction-1}} =& \int \frac{d^{4} l}{(2 \pi)^4}  V_{\alpha \beta, {\rm eff.}}^{\text{(1)spin-2}}(k_1,k_1',M) \frac{-i \mathcal{P}^{\alpha \beta}_{\;\;\;\gamma \delta}}{q^2} V^{(3)\gamma \delta }_{\lambda \psi \phi \epsilon} (l,-q) \frac{-i \mathcal{P}^{\phi \epsilon \rho \sigma }}{(l+q)^2} \frac{-i \mathcal{P}^{\lambda \psi \mu \nu }}{l^2} \\
            & \times V_{\mu \nu }^{\text{(1)spin-0}} (k_2,k_2-l,m) V_{\rho \sigma }^{\text{(1)spin-0}} (k_2-l,k_2',m) \frac{-i}{(l-k_2)^2 + m^2},
        \end{split}
\end{equation}
which yields,
\begin{equation}
\begin{split}
    \mathcal{M}_{\text{vertex-correction-1}} =& \Bigg[ \frac{46 \pi ^2 G^2 m^3 M^2 }{q}  + \frac{4 \pi ^2 G^2 m^2 M^3 }{q}  + 46 G^2 m^2 M^2  \ln q^2 \Bigg]  \epsilon_{\mu \nu}^{(a)}(k_1) \epsilon^{\star \,\mu \nu \, (a')}(k'_1) \\
    & + {\rm subleading~terms}.
\end{split}
\end{equation}
The corresponding correction to the Newton potential reads,
\begin{equation}
\begin{split}
    V_{\text{vertex-correction-1}} (r) =& \Bigg[ - \frac{23 G^2 m^2 M }{4 r^2} - \frac{G^2 m M^2 }{2 r^2} + \frac{23 G^2 m M }{4 \pi  r^3} \Bigg] \epsilon_{ij}^{(a)}(\vec{k}_1) \epsilon^{\star \, ij\, (a')}(\vec{k}'_1) \\
    & +  {\rm  subleading~terms}.
\end{split}
\end{equation}
%
\subsection{The vertex correction diagram-2}
\begin{figure}[h!]
\begin{center}
    \includegraphics[width=0.3\linewidth]{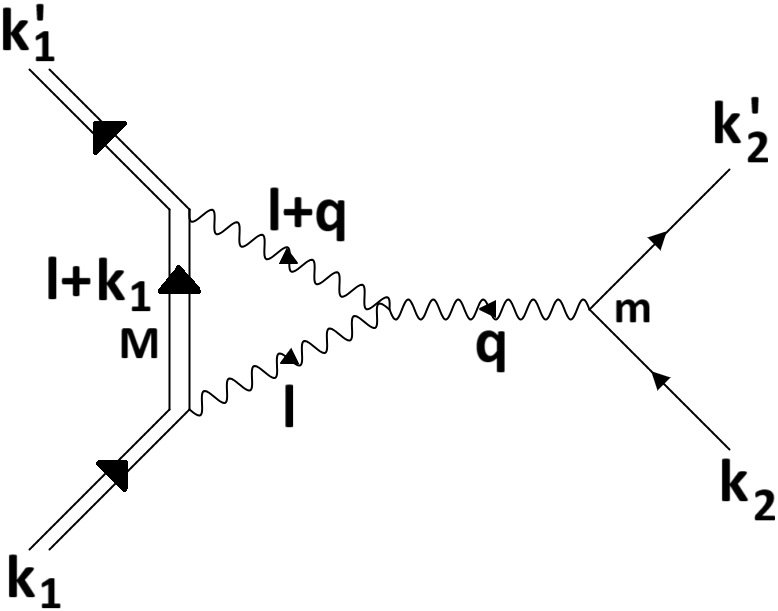} 
  \caption{\small \it The second vertex correction diagram for massive spin-2-spin-0 fields scattering.   }
   \label{fl7}
\end{center}
\end{figure}

The second vertex correction diagram is given by \ref{fl7}.
The Feynman amplitude reads,
\begin{equation}
    \begin{split}
        i \mathcal{M}_{\text{vertex-correction-2}} =&  \int\frac{d^4l}{(2\pi)^4}  V^{(1)\,\text{spin-2}}_{\mu  \nu \text{, eff.}}(k_1,l+k_1, M) \epsilon^\star _{\lambda' \eta '}(k_1') \frac{-i {\cal P}_{M}^{\lambda' \eta ' \epsilon ' \psi' }}{(l+k_1)^2+M^2}  V^{(1)\,\text{spin-2}}_{\alpha \beta  \text{, eff.}}(l+k_1,k'_1, M)  \\
        &\times \epsilon _{\epsilon ' \psi'}(k_1) \frac{- i \mathcal{P}^{\mu \nu  \rho \sigma}}{l^2} \  \frac{ - i \mathcal{P}^{\alpha \beta  \gamma \delta}}{(l+q)^2}  V^{(3)\phi \epsilon }_{\rho \sigma \gamma \delta} (-l,q) \ \frac{- i P_{\;\phi \epsilon \lambda \psi}}{q^2} V^{\lambda \psi }_{(1)\text{spin-0}} (k_2,k_2',m),
    \end{split}
\end{equation}
which yields, 
\begin{equation}
\begin{split}
    \mathcal{M}_{\text{vertex-correction-2}} =&  \Bigg[\frac{24 \pi ^2 G^2 m^3 M^2  }{q} - \frac{71 \pi ^2 G^2 m^2 M^3  }{q} - 71 G^2 m^2 M^2   \ln q^2 \Bigg]  \epsilon_{\mu \nu}^{(a)}(k_1) \epsilon^{\star \,\mu \nu \, (a')}(k'_1) \\
    & + {\rm  polarisation-dependent~ subleading~terms}.
\end{split}
\end{equation}
The correction to the gravitational potential reads,
\begin{equation}
\begin{split}
    V_{\text{vertex-correction-2}} (r) =& \Bigg[-\frac{3 G^2 m^2 M  }{r^2} +\frac{71 G^2 m M^2  }{8 r^2} - \frac{71 G^2 m M  }{8 \pi  r^3} \Bigg] \epsilon_{ij}^{(a)}(\vec{k}_1) \epsilon^{\star \, ij\, (a')}(\vec{k}'_1) \\
     & + {\rm   subleading~terms}.
\end{split}
\end{equation}
%

\subsection{ The vertex correction diagram-3}
\begin{figure}[h!]
\begin{center}
    \includegraphics[width=0.35\linewidth]{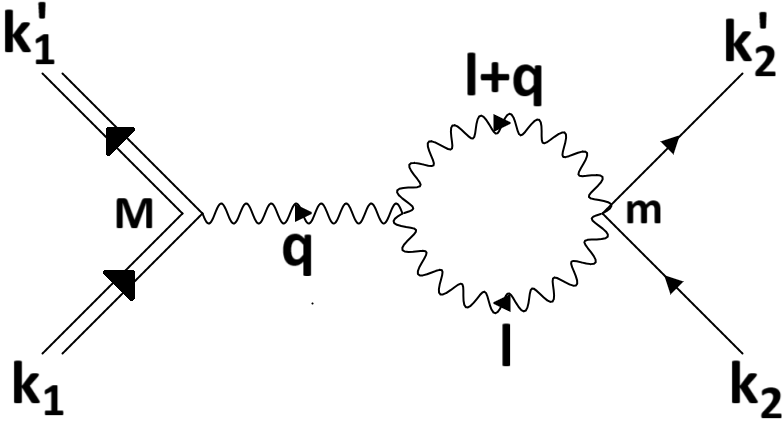} 
  \caption{\small \it The third vertex correction diagram for massive spin-2-spin-0 fields scattering.}
   \label{fl8}
\end{center}
\end{figure}

The third vertex correction diagram is given by \ref{fl8}.
The corresponding Feynman amplitude reads,
\begin{equation}
    \begin{split}
        i \mathcal{M}_{\text{vertex-correction-3}} = &\frac{1}{2!} \int \frac{d^{\;4} l}{(2 \pi)^4}  V_ {\alpha \beta}^{\:\text{(1)spin-2, eff. }}(k_1,k_1',M)   \frac{ -i \mathcal{P}^{\,\alpha \beta}_{\;\;\gamma \delta}}{(l+q)^2}  V^{(3)\gamma \delta}_{\mu \nu \rho \sigma} (l,-q)  \ \frac{- i \mathcal{P}^{ \rho \sigma \epsilon \phi}}{q^2} \frac{-i \mathcal{P}^{\mu \nu \lambda \psi}}{l^2} \\
        & \times V_ {\lambda \psi \epsilon \phi }^{\:\text{(2)spin-0}} (k_2,k_2',m),
    \end{split}
\end{equation}
which gives,
\begin{equation}
    \mathcal{M}_{\text{vertex-correction-3}} =  80 G^2 m^2 M^2  \ln q^2   \epsilon_{\mu \nu}^{(a)}(k_1) \epsilon^{\star \,\mu \nu \, (a')}(k'_1)  + {\rm  polarisation-dependent ~subleading~terms}
\end{equation} 
The corresponding potential reads,
\begin{equation}
     V_{\text{vertex-correction-3}} (r) =  \frac{10 G^2 m M  }{\pi  r^3}\epsilon_{ij}^{(a)}(\vec{k}_1) \epsilon^{\star \, ij\, (a')}(\vec{k}'_1) + {\rm  subleading~terms}. 
\end{equation}
%

\subsection{ The vertex correction diagram-4} 
%
\begin{figure}[h!]
\begin{center}
    \includegraphics[width=0.35\linewidth]{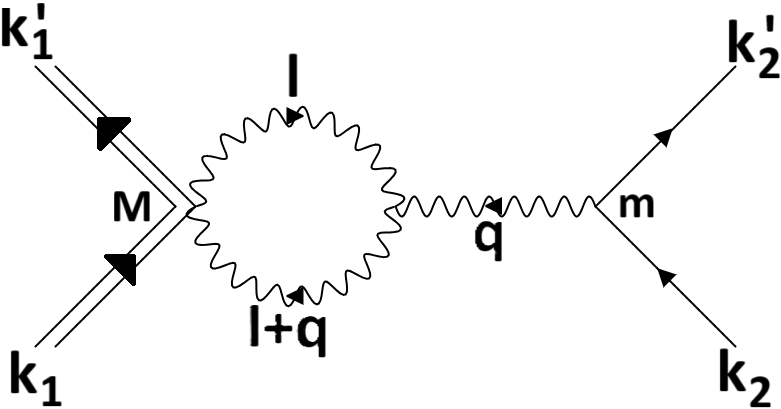} 
  \caption{\small \it The fourth vertex correction diagram for massive spin-2-spin-0 fields scattering.   }
   \label{fl9}
\end{center}
\end{figure}
\noindent The fourth and the last vertex correction diagram is given by \ref{fl9}. The Feynman amplitude reads,
\begin{equation}
    \begin{split}
       i \mathcal{M}_{\text{vertex-correction-4}} =& \frac{1}{2!} \int \frac{d^{\;4} l}{(2 \pi)^4} V_{(2)\text{spin-2}}^{\rho\sigma\mu\nu}(k_1,k'_1,M ) V_{(1)\text{spin-0}}^{\lambda\kappa}(k_2,k'_2,m) V_{(3)}^{\alpha\beta\gamma\delta\epsilon\phi}(-l,q) \frac{-i{\cal P}_{\mu\nu\gamma\delta}}{l^2} \frac{-i{\cal P}_{\rho\sigma\alpha\beta}}{(l+q)^2} \frac{-i{\cal P}_{\lambda\kappa\epsilon\phi}}{q^2},
    \end{split}
\end{equation}
which equals,
\begin{equation}
    \mathcal{M}_{\text{vertex-correction-4}} = \frac{21}{4} G^2 m^2 M^2 \ln q^2 \  \epsilon_{\mu \nu}^{(a)} (k_1)\epsilon^{\star \,\mu \nu \, (a')}(k'_1) + {\rm  subleading ~terms}.
\end{equation} 
The corresponding correction to the  gravitational potential reads,
\begin{equation}
     V_{\text{vertex-correction-4}} (r) = \frac{21 G^2 m M }{64 \pi r^3} \epsilon_{ij}^{(a)}(\vec{k}_1) \epsilon^{\star \, ij\, (a')}(\vec{k}'_1) + \ {\rm  subleading~terms}. 
\end{equation}
%

\subsection{The vacuum polarisation and the ghost diagrams}

\begin{figure}[h!]
\begin{center}
    \includegraphics[width=0.68\linewidth]{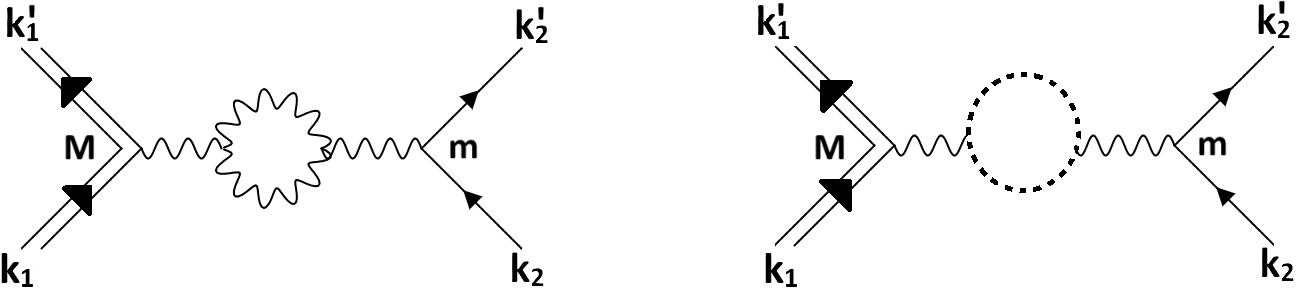} 
  \caption{\small \it The vacuum polarisation plus ghost diagrams for massive spin-2-spin-0 fields scattering.   }
   \label{fl10}
\end{center}
\end{figure}

\begin{figure}[h!]
\begin{center}
    \includegraphics[width=0.5\linewidth]{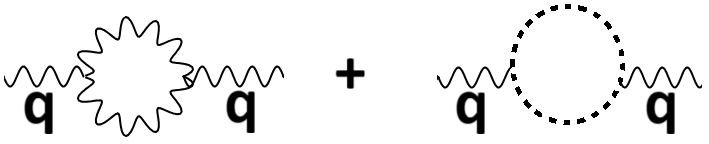} 
  \caption{\small \it Graviton self energy from the vacuum polarisation diagram along with the ghost contribution.   }
   \label{fl11}
\end{center}
\end{figure}

Finally, we come to the contribution from the vacuum polarisation plus the ghost loop, \ref{fl10}.  The ghost loop is added to compensate for the gauge non-invariant terms originating from the graviton self energy, \ref{fl11}. The total gauge invariant contribution denoted as $\Pi_{\lambda \xi \mu \nu} (q)$, has been evaluated long ago in~\cite{tHooft:1974toh,  vanDam:1970vg},
\begin{equation}
\begin{split}
            \Pi_{\alpha \beta \gamma \delta } =& -\frac{2G}{\pi}\ln q^2 \Big[\frac{21}{120}q^4 I_{\alpha \beta \gamma \delta} + \frac{23}{120}q^4 \eta_{\alpha \beta}\eta_{\gamma \delta}- \frac{23}{120}q^2 (\eta_{\alpha \beta}q_{\gamma} q_{\delta} + \eta_{\gamma \delta} q_{\alpha} q_{\beta}) - \frac{21}{240}q^2 (q_{\alpha} q_ {\delta}\eta_{\beta \gamma} + q_{\beta} q_ {\delta}\eta_{\alpha \gamma} \\
            & + q_{\alpha} q_ {\gamma}\eta_{\beta \delta} + q_{\beta} q_ {\gamma}\eta_{\alpha \delta}) + \frac{11}{30}q_{\alpha} q_{\beta} q_{\gamma}q_{\delta}\Big]
        \end{split}
        \label{vpol}
\end{equation}
We have the Feynman amplitude,  
\begin{equation}
    i\mathcal{M}_{\text{vac-pol}} = V^{\text{(1)spin-2, eff.}}_{\rho \sigma} (k_1,k_1',M) \ \dfrac{ - i \mathcal{P}^{\rho \sigma \lambda \xi }}{q^2} \Pi_{\lambda \xi \mu \nu} (q) \ \dfrac{- i \mathcal{P}^{\mu \nu \gamma \delta}}{q^2}   V^{\text{(1)spin -0}}_{\gamma \delta} (k_2,k_2',m),
\end{equation}  
which equals after substituting Eq.~\ref{vpol},
\begin{equation}
    \mathcal{M}_{\text{Ghost}} = - \frac{172}{15} G^2 m^2 M^2 \ln q^2  \epsilon_{\mu \nu}^{(a)} (k_1) \epsilon^{\star \,\mu \nu \, (a')}(k'_1)+{\rm subleading~terms} .
\end{equation}
The corresponding potential reads,
\begin{equation}
     V_{\text{vac-pol}}(r) = -  \frac{43 M m G^2}{30\pi r^3} \epsilon_{ij}^{(a)}(\vec{k}_1) \epsilon^{\star \, ij\, (a')}(\vec{k}'_1) + \ {\rm   subleading~terms}. 
\end{equation}
%

\subsection{The final quantum corrected Newtonian potential }
Combining everything now, we obtain the leading part of the   gravitational potential up to ${\cal O}(G^2)$, between a massive spin-2 and spin-0 particle,
\begin{equation}
    \begin{split}
         V(r) =&   -\frac{G M m}{r} \Bigg[1 +\frac{ G (171m-110M)  }{24 r c^2}+\frac{577 G \hbar }{80 \pi  r^2 c^3}\Bigg]  \epsilon_{ij}^{(a)}(\vec{k}_1) \epsilon^{\star \, ij\, (a')}(\vec{k}'_1) + \ {\rm ~subleading~terms}.  
    \end{split}
    \label{total}
\end{equation}
This is one of the main results of this paper. Note that we have put back $\hbar$ in the last term. This is to emphasise that this term has purely quantum origin -- i.e. from the $\ln q^2$ term coming from the loops.
\section{Discussion}{\label{S5}
Let us summarise our results now. In this paper we have computed the 2-2 scattering between a massive spin-2 Fierz-Pauli field with various other spin fields, and from the non-relativistic limit of the scattering amplitudes, we have computed the corresponding two body gravitational potentials at the leading (${\cal O}(G)$) and next to the leading (${\cal O}(G^2)$) orders. The massive spin-2 field does {\it not} stand for any massive gravity here, and gravity is taken to be simply general relativity with the usual massless graviton as its quantum. We have taken the Fierz-Pauli massive spin-2 field as just a field theory with higher spin, minimally coupled to gravity, e.g.~\cite{Bekenstein:1972ky}. At the leading order (${\cal O}(G)$), we have computed the gravitational potential for its scattering with massive scalar, fermion, and Proca fields, \ref{S3}. The Newton potential, along with the spin and polarisation dependent non-spherical terms, have been explicitly computed.  For the massive spin-2 - massive scalar scattering, we have further extended the above computations to the next to the leading order (${\cal O}(G^2)$), \ref{S4}.  For this case,  we have explicitly demonstrated  the polarisation independent spherical parts only. The present computations could be thought of as one of the ongoing attempts to understand the gravitational potential in the context of a higher spin field theory.

As a step forward, perhaps it will be interesting to understand the quantum corrections to Newton's potential in viable massive gravity theories, in which the vDVZ discontinuity can be taken care of via some Vainshtein screening mechanism at scales small compared to the Hubble horizon~\cite{deRham:2014zqa}. We hope  to investigate this issue in our future work.

\bigskip

\noindent
{\bf\large Acknowledgements :} ASM would like to thank V.~Nath and A. K. Naskar for useful discussions and help. He also acknowledges N.~E.~J.~Bjerrum-Bohr for some useful communications. The authors would like to sincerely acknowledge anonymous referee for various  useful  critical comments. 

\bigskip
\section{Appendix} 
\appendix
\labelformat{section}{Appendix #1} 
\section{Various interaction terms}
\label{A0}  
\subsection{Massive spin-0 field}\label{A01}

The action for a massive  scalar field  is given by,
\begin{equation}
S^{\text{\:spin-0}} = -\frac12\int \sqrt{-g} \left[g^{\mu\nu}(\p_{\mu} \phi)(\p_{\nu} \phi) + m^2 \phi^2 \right]
\end{equation} 
The free as well as the scalar-graviton interaction terms up to ${\cal O}(\kappa^2)$ read (using Eq.~\ref{qg17}),
\begin{equation}
\begin{split}
    &  S^{\text{\:spin-0}}(\kappa^0)= -\frac12 \int d^4 x \Big[(\p_{\mu}\phi)(\p^{\mu}\phi)+m^2 \phi^2 \Big]\\ 
&    S^{\text{\:spin-0}}(\kappa^1) =\frac{\kappa}{2} \int d^4 x h^{\mu\nu}\left[(\p_{\mu}\phi)(\p_{\nu}\phi) -\frac12 \eta_{\mu\nu}\left( (\p \phi)^2+m^2\phi^2 \right)\right] \\
 &   S^{\text{\:spin-0}}(\kappa^2) = -\frac{\kappa^2}{2}\int d^d x \left[ \left( h^{\mu}{}_{\alpha} h^{\alpha \nu} -\frac12 h h^{\mu\nu} +\frac18 h^2 \eta^{\mu\nu} -\frac14 h_{\alpha\beta} h^{\alpha\beta} \eta^{\mu\nu} \right)(\p_{\mu}\phi) (\p_{\nu}\phi) +m^2 \left(\frac{h^2}{8}-\frac14 h_{\mu\nu}h^{\mu\nu} \right)\phi^2\right]\\
\end{split}
\end{equation}
%

\subsection{Massive spin-1 field}\label{A}

For a massive spin-1 field we have,
\begin{equation}
    S^{\text{\,spin-1}} = \int \sqrt{-g}\Big[-\frac 14 F_{\mu\nu}F^{\mu\nu} -\frac12 m^2_1 A_{\mu} A^{\mu} \Big]
\label{qg0'}
\end{equation}
where $m_v$ is the mass of the spin-1 particle and, $F_{\mu\nu}= \p_{\mu}A_{\nu}- \p_{\nu}A_{\mu}$. The free as well as the  two spin-1 - one  graviton interaction terms  read
\begin{equation}
\begin{split}
    \label{A04a}
 &   S^{\text{\:spin-1}}(\kappa^0)= -\int d^4 x \left[ \frac14 \eta^{\mu \lambda}\eta^{\nu\rho} F_{\mu\nu}F_{\lambda \rho}+\frac12 m_v^2\eta^{\mu\nu}A_{\mu}A_{\nu}\right]
\nonumber\\
&    S^{\text{\:spin-1}}(\kappa^1) =\frac{\kappa}{2}\int d^4 x \left[ \left(\eta^{\mu\lambda}h^{\nu\rho}-\frac14 \eta_{\alpha\beta}\eta^{\mu\lambda}\eta^{\nu \rho}h^{\alpha\beta} \right) F_{\mu\nu}F_{\lambda \rho}+ m_v^2\left(h^{\mu\nu}-\frac12 \eta_{\alpha \beta}\eta^{\mu\nu} h^{\alpha \beta} \right)A_{\mu}A_{\nu} \right].
\end{split}
\end{equation}
This will be sufficient for our present purpose with the massive spin-1 field.

\subsection{Massive spin-2 field}

Let us now consider the Fierz-Pauli massive spin-2 field. The Lagrangian is given in Eq.~\ref{ml1} from where the action can be easily written. We have  up to ${\cal O}(\kappa^2)$,
\begin{equation}
    \begin{split}
        S^{\text{\:spin-2}}(\kappa^0) =& \int d^4 x \Big[- \frac12 (\p_{\lambda}H^{\mu\nu})(\p^{\lambda}H_{\mu\nu})+(\p^{\lambda}H_{\mu\lambda})(\p_{\alpha}H^{\mu\alpha}) -(\p_{\lambda}H)(\p_{\mu}H^{\mu\lambda})+ \frac12 (\p_{\lambda}H)(\p^{\lambda}H) \\
        & -\frac12 M^2 (H_{\mu\nu}H^{\mu\nu}-H^2)\Big]
    \label{A01}
    \end{split}
\end{equation}
\begin{equation}
    \begin{split}
        S^{\text{\:spin-2}}(\kappa^1) =& -\frac{\kappa}{2} \int d^4 x \Big[ \Big(  2\eta^{\lambda \rho}\eta_{\nu\beta}h_{\mu\alpha} - \eta_{\mu\alpha}\eta_{\nu\beta}h^{\lambda\rho}+\frac{h}{2}\eta^{\lambda\rho}\eta_{\mu\alpha}\eta_{\nu\beta}\Big)(\p_{\lambda}H^{\mu\nu})(\p_{\rho}H^{\alpha\beta})\\
        &+ 2\eta^{\lambda \rho}\eta_{\nu\beta}(\p_{\rho}h_{\mu\alpha}+ \p_{\alpha}h_{\rho\mu}-\p_{\mu}h_{\rho\alpha})H^{\alpha\beta}(\p_{\lambda}H^{\mu\nu}) + M^2\Big(2\eta_{\mu\alpha}h_{\nu\beta}+ \frac{h}{2}\eta_{\mu\alpha}\eta_{\nu\beta}\Big)H^{\mu\nu}H^{\alpha\beta} \Big]
\label{A02}
    \end{split}
\end{equation}
and,
\begin{eqnarray}
    \begin{split}
        S^{\text{\:spin-2}}(\kappa^2) =& \kappa^2\int d^4 x \Bigg[ -\frac12 \Bigg(\eta^{\lambda \rho}h_{\mu\alpha}h_{\nu\beta}-2\eta_{\nu\beta}h_{\mu\alpha}h^{\lambda\rho}+ h^{\lambda}{}_{\beta'}h^{\beta'\rho}\eta_{\mu\alpha}\eta_{\nu\beta}+ hh_{\mu\alpha}\eta^{\lambda\rho}\eta_{\nu\beta}-\frac12h h^{\lambda \rho}\eta_{\mu\alpha}\eta_{\nu\beta}\\
        & +\frac14 \eta^{\lambda \rho}\Big(\frac{h^2}{2}- h_{\alpha'\alpha''}h^{\alpha'\alpha''} \Big)  \Bigg)(\p_{\lambda}H^{\mu\nu})(\p_{\rho}H^{\alpha\beta})-\frac12\Big(2\eta^{\lambda \rho}\eta_{\nu\beta}h_{\mu\alpha}- \eta_{\mu\alpha}\eta_{\nu\beta}h^{\lambda \rho}+\frac{h}{2}\eta^{\lambda\rho}\eta_{\mu\alpha} h_{\nu\beta} \Big) \\
        & (\p_{\lambda}H^{\mu\nu})\Big(\eta^{\alpha\alpha'}(\p_{\rho}h_{\rho'\alpha'}+\p_{\rho'}h_{\rho\alpha'}-\p_{\alpha'}h_{\rho\rho'} )H^{\rho'\beta}  + \eta^{\beta\alpha'}(\p_{\rho}h_{\rho'\beta'}+\p_{\rho'}h_{\rho\beta'}-\p_{\beta'}h_{\rho\rho'} )H^{\rho'\alpha}      \Big)\\
        & +\frac14 \Big((2\p_{\lambda}h_{\mu\nu}-\p_{\mu}h_{\lambda \nu})H^{\nu\lambda}+ \eta_{\rho \mu}\eta^{\lambda\lambda'}\p_{\nu} h_{\lambda \lambda'}H^{\nu\rho}  \Big) \Big(\eta^{\mu\mu'}(2\p_{\alpha}h_{\mu'\beta}-\p_{\mu'}h_{\alpha\beta})H^{\alpha \beta}+ \eta^{\alpha \alpha'} \p_{\beta} h_{\alpha\alpha'}H^{\mu\beta} \Big) \\
        & +\frac12 \Big((\p_{\mu}h_{\alpha\beta})(\p^{\mu}h_{\lambda\rho})H^{\alpha\beta}H^{\lambda\rho} + 2(\p_{\mu}h_{\alpha\beta})(\p^{\mu}H^{\lambda \rho})h_{\lambda \rho}H^{\alpha \beta}+ h_{\alpha\beta}h_{\lambda \rho}(\p_{\mu}H^{\alpha\beta})(\p^{\mu}H^{\lambda \rho}) \Big) \\
        & - \frac12 \Big( (\p_{\lambda} h_{\alpha \beta})H^{\alpha \beta}+ h_{\alpha \beta}(\p_{\lambda}H^{\alpha \beta})\Big)\Big(\eta^{\mu\mu'}(\p_{\rho}h_{\mu\mu'} )H^{\rho\lambda}+ \eta^{\lambda\lambda'}(2\p_{\mu}h_{\rho\lambda'}- \p_{\lambda'}h_{\mu\rho}) H^{\mu\rho}\Big)\\
        & - \frac12 M^2 \Big(h_{\mu\lambda}h_{\nu\rho}- h_{\mu\nu}h_{\lambda \rho}+ h \eta_{\mu\lambda} h_{\nu\rho}+ \frac14 \Big(\frac{h^2}{2}-h_{\alpha\beta}h^{\alpha\beta} \Big) \eta_{\mu\lambda}\eta_{\nu\rho}\Big)H^{\mu\nu}H^{\lambda \rho}\Bigg].
\label{A03}
    \end{split}
\end{eqnarray}
Note that the terms containing derivatives of the graviton field will carry its momentum explicitly. Such terms will always make a subleading (comapared to $r^{-3}$) or even no contribution to the long range gravitational potential.

%
\section{List of essential integrals}
\label{B}  
We list below some integrals useful for the purpose of evaluating the gravitational potential, originating from the parts non-analytic in the transfer momentum, $q^2$~\cite{Bjerrum-Bohr:2002gqz}. We will display below only those non-analytic parts.
\begin{eqnarray}
 &&       \int \dfrac{d^d l}{(2 \pi)^d} \dfrac{1}{l^2 (l+q)^2} =  -   \dfrac{i }{16 \pi^2} \ln q^2; \qquad   \int \dfrac{d^d l}{(2 \pi)^d} \dfrac{l_{\mu}}{l^2 (l+q)^2} = \dfrac{i q_{\mu}}{32 \pi^2} \ln q^2 \nonumber\\ &&
        \int \dfrac{d^d l}{(2 \pi)^d} \dfrac{l_{\mu}l_{\nu}}{l^2 (l+q)^2} = - \dfrac{i \ln q^2}{64 \pi^2} \Big[\dfrac{4}{3} q_{\mu} q_{\nu} - \dfrac{1}{3} q^2 \eta_{\mu \nu}\Big]\nonumber\\ &&
        \int \dfrac{d^d l}{(2 \pi)^d} \dfrac{1}{l^2 (l+q)^2 ((l+k)^2 - m^2)} = -\dfrac{i}{32 \pi^2  m^2} \Bigg[ \ln q^2 + \dfrac{\pi^2m}{q}  \Bigg]
\end{eqnarray}
\begin{eqnarray}
        \int \dfrac{d^d l}{(2 \pi)^d} \dfrac{l_{\mu}}{l^2 (l+q)^2 ((l+k)^2 - m^2)} = \dfrac{i}{32 \pi^2 m^2} \Big[ \Big\{\Big(-1 -  \dfrac{q^2}{2m^2}\Big)  \ln q^2 - \dfrac{1}{4} \dfrac{q^2}{m^2}  \dfrac{\pi^2 m}{q} \Big\}  k_{\mu} + \Big\{ \ln q^2 + \dfrac{\pi^2 m}{2 q} \Big\}q_{\mu}  \Big]
\end{eqnarray}
\begin{equation}
    \begin{split}
        \int \dfrac{d^d l}{(2 \pi)^d} \dfrac{l_{\mu} l_{\nu}}{l^2 (l+q)^2 ((l+k)^2 - m^2)} =& \dfrac{i}{32 \pi^2 m^2} \Big[ q_{\mu} q_{\nu} \Big( - \ln q^2 -\dfrac{3}{8} \dfrac{\pi^2 m}{q}  \Big) + k_{\mu} k_{\nu} \Big( - \dfrac{1}{2} \dfrac{q^2}{m^2} \ln q^2 - \dfrac{1}{8} \dfrac{q^2}{m^2} \dfrac{\pi^2 m}{q}  \Big)  \\
        &+ (q_{\mu} k_{\nu} + q_{\nu} k_{\mu}) \Big\{ \Big( \dfrac{1}{2} + \dfrac{1}{2} \dfrac{q^2}{m^2}  \Big)\ln q^2 + \dfrac{3 \pi^2 m q}{16}   \Big\} + q^2 \eta_{\mu \nu} \Big( \dfrac{1}{4} \ln q^2 +   \dfrac{\pi^2 m}{8 q}  \Big)  \Big]
    \end{split}
\end{equation}
\begin{eqnarray}
\label{A}
&& \int\frac{d^dl}{(2\pi)^d}
\frac{1}{l^2(l+q)^2((l+k_1)^2-m_1^2)((l-k_2)^2-m_2^2)} =  \dfrac{i}{16\pi^2m_1 m_2 q^2}\Big(1-\dfrac{w}{3m_1m_2}\Big)\ln q^2\nonumber\\
&& \int\frac{d^dl}{(2\pi)^d}
\frac{1}{l^2(l+q)^2((l+k_1)^2-m_1^2)((l+k'_2)^2-m_2^2)} = \dfrac{i}{16\pi^2m_1 m_2
q^2}\Big(-1+\dfrac{W}{3m_1m_2}\Big)\ln q^2
\end{eqnarray}
where,  $w = -k_1\cdot k_2-m_1m_2$ and $W = -k_1\cdot k'_2 - m_1m_2$. 
%

  
\end{document}